\documentclass[aps,prx,twocolumn,superscriptaddress]{revtex4-2}

\usepackage{hyperref}
\hypersetup{
  colorlinks,
  citecolor=blue,
  linkcolor=red}
\usepackage{graphicx}
\usepackage{amsmath,amssymb,latexsym}
\usepackage{xcolor}
\usepackage{ulem}
\usepackage{array}
\newcommand{\be}{\begin{equation}}
\newcommand{\ee}{\end{equation}}
\newcommand{\beq}{\begin{eqnarray}}
\newcommand{\eeq}{\end{eqnarray}}
\newcommand{\bee}{\begin{equation*}}
\newcommand{\eee}{\end{equation*}}
\newcommand{\beqq}{\begin{eqnarray*}}
\newcommand{\eeqq}{\end{eqnarray*}}
\newcommand{\w}{\omega}
\newcommand{\W}{\Omega}
\newcommand{\g}{\gamma}
\newcommand{\G}{\Gamma}

\newcommand{\s}{\sigma}

\newcommand{\ket}[1]{\left| #1 \right\rangle}
\newcommand{\bra}[1]{\left\langle #1 \right|}

\newcommand{\sbrac}[1]{\left\langle #1 \right\rangle}
\newcommand{\dbrac}[1]{\left\langle \! \left\langle #1 \right\rangle \! \right\rangle}
\newcommand{\sdbrac}[1]{\langle \! \langle #1 \rangle \! \rangle}
\newcommand{\TC}{{\cal T}_{\cal C}}
\newcommand{\SC}{{\cal S}_{\cal C}}
\newcommand{\Se}{{\cal S}^{\rm eff}_{\cal C}}
\newcommand{\T}{{\cal T}}
\newcommand{\cur}[1]{\left\{ #1 \right\}}
\newcommand{\squ}[1]{\left[ #1 \right]}
\newcommand{\pare}[1]{\left( #1 \right)}
\newcommand{\ih}{-\frac{i}{\hbar}}

\newcommand{\hide}[1]{}

\newcommand{\eq}[1]{Eq.\,(\ref{#1})}
\newcommand{\eqs}[1]{Eqs.\,(\ref{#1})}

\newcommand{\fig}[1]{Fig.\,\ref{#1}}

\newcommand{\tab}[1]{Table\,\ref{#1}}
\graphicspath{{Figures/}}

\begin{document}

\title{Superradiance and Subradiance in Dense Atomic Gases: An Integrated Method}

\author{Hanzhen Ma}
\affiliation{Department of Physics, University of Connecticut, Storrs, Connecticut 06269, USA}
\affiliation{Department of Physics, Harvard University, Cambridge, Massachusetts 02138, USA}
\author{Oriol Rubies-Bigorda}
\affiliation{Department of Physics, Harvard University, Cambridge, Massachusetts 02138, USA}
\affiliation{Physics Department, Massachusetts Institute of Technology, Cambridge, Massachusetts 02139, USA}
\author{Susanne F. Yelin}
\affiliation{Department of Physics, Harvard University, Cambridge, Massachusetts 02138, USA}

\date{\today}

\begin{abstract}
When atoms are coupled to a common electromagnetic environment, the exchange of photons through dipole-dipole interactions leads to the emergence of cooperative effects. As a particular example, {\it superradiance} arises from spontaneous emission when this exchange leads to constructive interference of the emitted photons. Here, we introduce an integrated method for studying cooperative radiation in many-body systems. This method, which allows to study extended systems with arbitrarily large number of particles, can be formulated by an effective, nonlinear, two-atom master equation that describes the dynamics using a closed form which treats single- and many-body terms on an equal footing. We apply this method to a homogeneous gas of initially inverted two-level atoms, and demonstrate the appearance of both superradiance and subradiance, identifying a many-body coherence term as the source of these cooperative effects. We describe the many-body induced broadening -- which is analytically found to scale with the optical depth of the system -- and light shifts, and distinguish spontaneous effects from induced ones. In addition, we theoretically predict the time-dependence of subradiance, and the phase change of the radiated field during the cooperative decay.
\end{abstract}

\maketitle

\section{Introduction}

\begin{figure}
	\includegraphics[width=\linewidth]{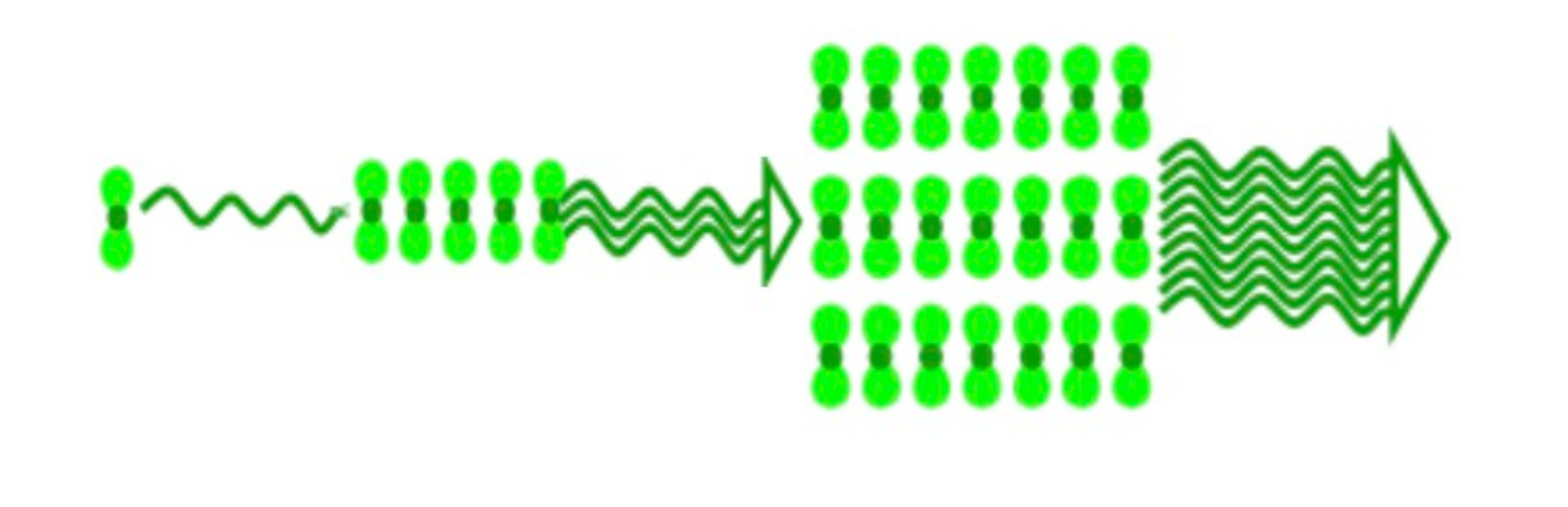}
	\caption{Visualization of superradiance. A photon emitted by one particle stimulates the close-by neighbors in turn, resulting in enhanced radiation.}
	\label{fig:atom_pic}
\end{figure}

\begin{figure}
        \includegraphics[width=0.7\linewidth]{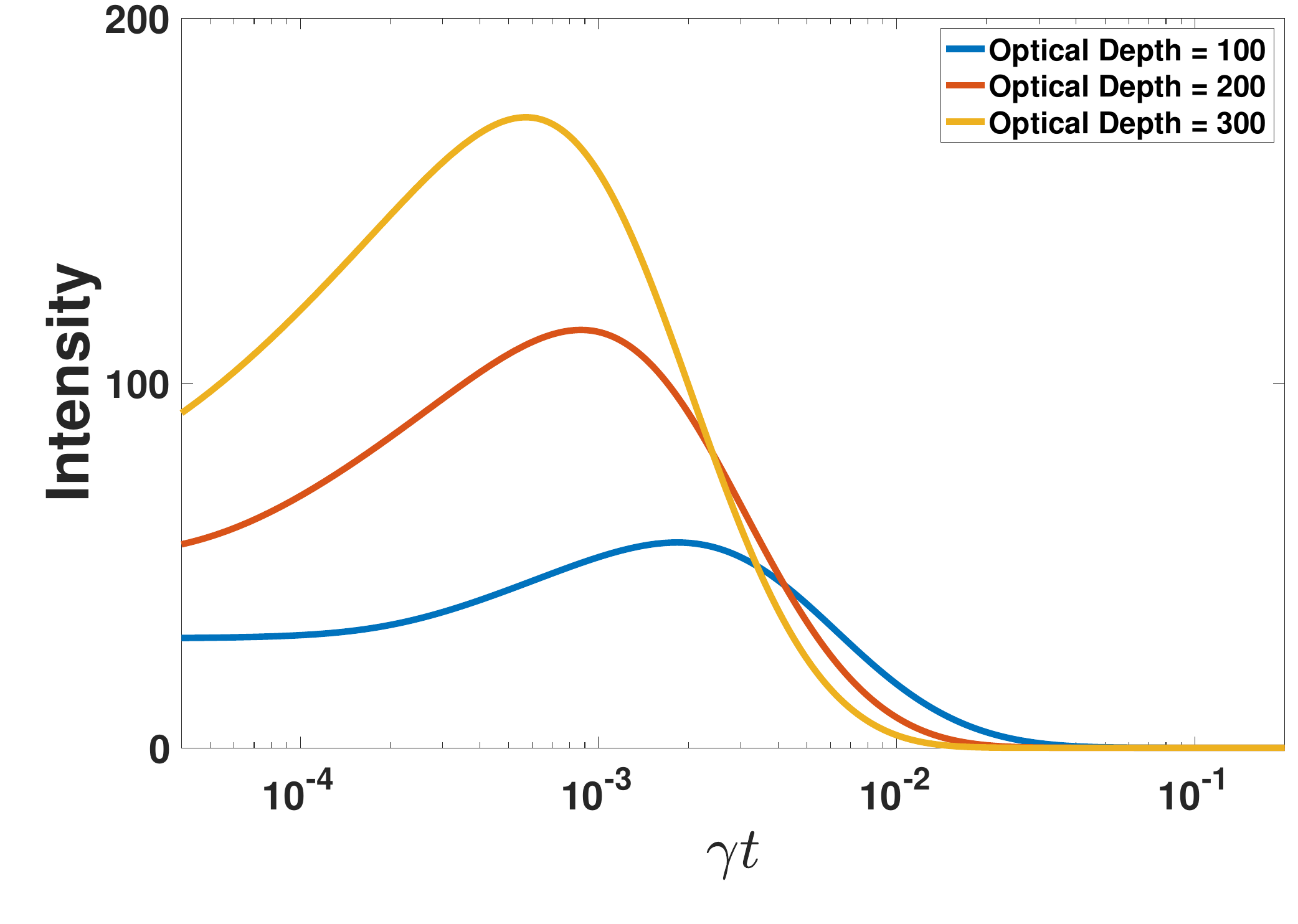}
	\caption{Radiated intensity (in arbitrary units) for superradiance with different optical depths of the sample, plotted with a dimensionless time scale, where $\g$ is the vacuum spontaneous decay rate.}
	\label{fig:intensity}
\end{figure}

Cooperative phenomena in many-body radiative systems result from the build-up of correlation among the radiators \cite{dicke1954,gross1982}. The dipole-dipole interaction mediated by a shared electromagnetic field gives rise to the all-to-all coupling of the radiators, leading to a dramatically different behavior compared to that of independent ones. As first presented in the pioneering work by Dicke \cite{dicke1954}, {\it superradiance} results from the fact that a spontaneously emitted photon from one particle can stimulate emission from close-by neighbors (\fig{fig:atom_pic}) \cite{rehler1971,feld1976,Lehmberg_Paper1,Lehmberg_Paper2}. The emitted photons are phase coherent and therefore lead to an enhancement of radiated intensity (\fig{fig:intensity}). {\it Subradiance}, on the contrary, results from the destructive interference between the emitted photons which suppresses radiation. These cooperative effects have been experimentally observed in various physical platforms, ranging from dense disordered atomic systems \cite{Experiment_dense_1,Experiment_dense_2,Experiment_dense_3,Experiment_dense_4,Experiment_dense_5,Experiment_dense_6,Experiment_dense_7,subradiance_densecloud} and ensembles of atoms in a cavity \cite{experiment_cavity_1,experiment_cavity_2,experiment_cavity_3} to condensed matter systems such as two-dimensional materials \cite{experiment_2dmat} and quantum dots \cite{experiment_quantumdots_1,experiment_quantumdots_2}. While superradiance has been extensively studied due to its applications in laser techniques \cite{meiser2009,bohnet2012} and the generation of spin-squeezing and entanglement \cite{elie2014,spin_squeezing1,entanglement1,entanglement2,entanglement3}, subradiant atomic modes offer promising avenues for sensing \cite{sensing_1,sensing_2}, metrology \cite{metrology} and light storage and retrieval \cite{storage_oriol,storage_ruostekoski}.

The purpose of this article is to develop a formalism that approximates large-sample superradiance while incorporating full quantum effects. Standard approaches, such as the coupled-dipole model, typically rely on tracing out the electromagnetic degrees of freedom and deriving a master equation for the $N$ interacting atoms. However, this model features a Hilbert space whose size increases exponentially as $2^N$, posing significant challenges for both theoretical approaches and numerical calculations. In this work, we demonstrate that our method differs from the conventional coupled-dipole model in several key aspects, as detailed in Appendices \ref{app:A} and \ref{app:B}. Efforts have been made to reduce the complexity of the problem, including the case where all atoms are assumed to be within a volume of side length much smaller than the transition wavelength such that the system is described by the symmetric Dicke states
\cite{dicke1954,gross1982,dicke_superradiance1,dicke_superradiance2,dicke_superradiance3}. Other methods to describe collective effects rely on classical simulations \cite{javanainen2014,lee2016}, restricting the system to singly-excited states \cite{single_excitation_scully,single_excitation_1,single_excitation_2,single_excitation_3,single_excitation_4,single_excitation_5,single_excitation_6}, considering ensembles with a small number of particles such that quantum Monte-Carlo wave function techniques can be applied \cite{few_atoms_charmichael_2,few_atoms_charmichael,ana2020_fewatoms}, or neglecting high-order correlations via a cumulant expansion of the equations of motion ~\cite{cumulant_expansion_3,cumulant_expansion_4,cumulant_expansion_5,cumulant_expansion_6,cumulant_expansion_7}. The methods above reduce the full Hilbert space to a low-dimensional, relevant subspace and therefore make the problem feasible to solve, at the expense of compromising its applicability to systems with certain prerequisites. Recent studies exploit the system’s behavior at the initial time to predict the universal properties of superradiance for inverted ensembles with large number of atoms \cite{masson2021,robicheaux2021}.
While these formalisms do not require to solve the full time evolution of the system, they provide no information about the magnitude of the superradiant burst or the late time dynamics of the atomic ensemble.

In addition, there has been a large number of experimental demonstrations recently, such as those in Refs.~\cite{Das20,Gold22,ferioli21,weiss21,Experiment_dense_7}, which can, as an overall sample of experimental results, only be confirmed theoretically by a comprehensive treatment of large-scale and long-time theoretical treatment.

In this article, we propose an integrated method to study correlated systems that exhibit cooperative phenomena such as superradiance and subradiance. Our goal is to obtain an effective description of a system with a large particle number, whose degrees of freedom are reduced to a solvable level. Additionally, this description does still preserve the many-body nature of the problem arising from quantum correlations. For that, we follow the idea of a mean-field plus second-order correlations approach by tracing out the electromagnetic field and all but two atoms. As a result, we obtain the dynamics of the two remaining probe atoms, which reveal the cooperative behavior of arbitrarily large systems. As opposed to other treatments, the present formalism also describes the analytical dependencies of the emission properties on the parameters of the system, such as the optical depth, the particle density and the magnitude of the external driving field. 

\begin{figure}
	\includegraphics[width=\linewidth]{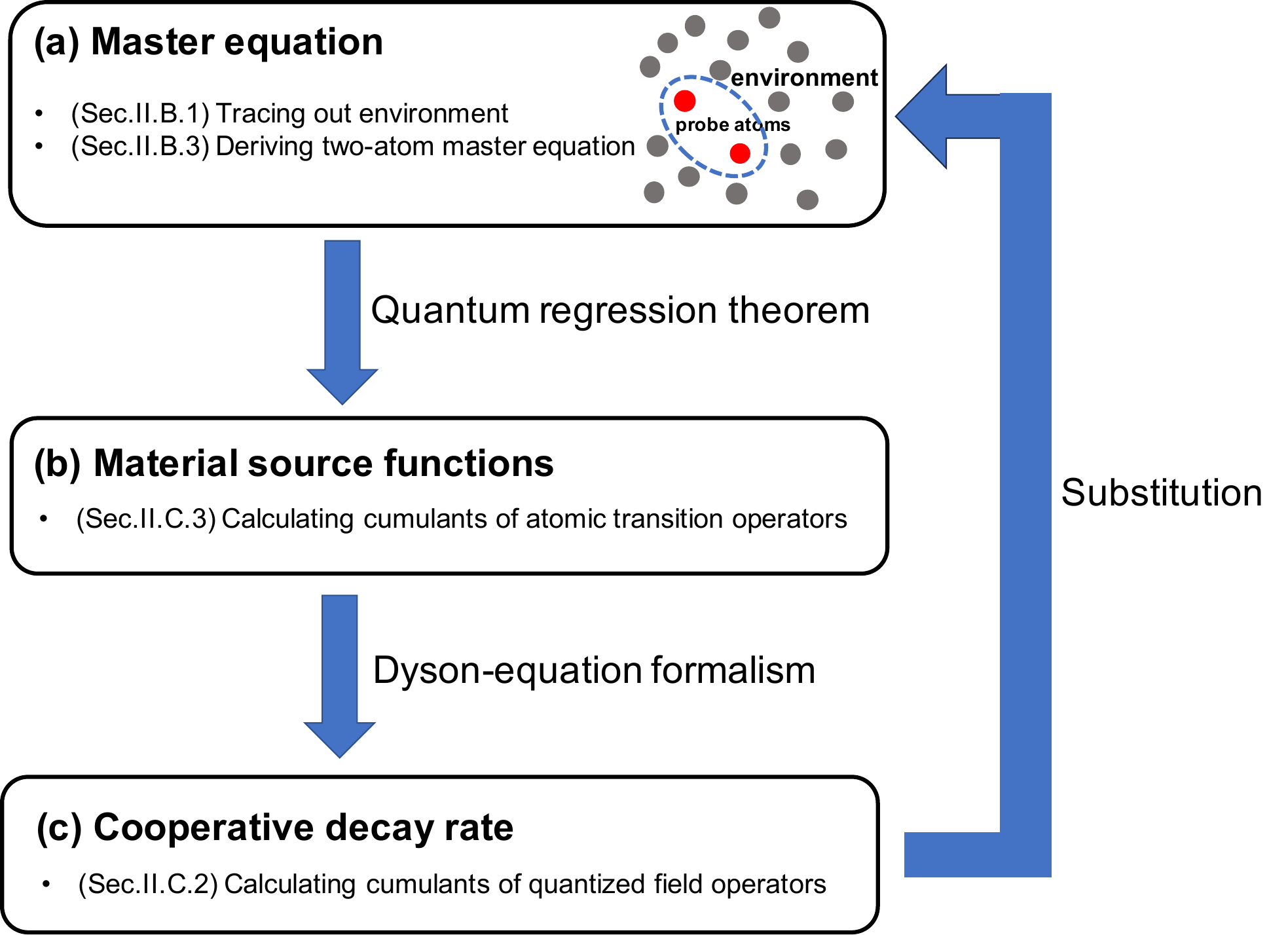}
	\caption{Schematic diagram of the closed-form formalism, showing the procedure of the integrated method. Related sections in this paper are indicated in parentheses.}
	\label{fig:closed_form}
\end{figure}

Our formalism can be briefly outlined as follows: (i) we start with the full Hamiltonian of the system (\eq{eq:full_H}), which describes the atoms interacting with a shared quantized field and a classical driving field. The Hamiltonian is then divided into two parts: an interaction term $V$, which consists of the interaction of two arbitrarily chosen probe atoms with the electromagnetic field, and $H_0$, which includes the interaction of the field with all the remaining atoms. (ii) We trace out the quantum field and the non-probe atoms, such that the dynamics effectively results from an atom-atom interaction and the degrees of freedom are dramatically reduced. This effective atom-atom interaction takes into account multiple scattering to all orders. (iii) We then develop a self-consistent formalism (see \fig{fig:closed_form}) in which the effective description of the full atomic ensemble is given by an average over all possible choices of probe atoms. (iv) This results in one main result of the paper --- a nonlinear two-atom master equation of Lindblad form that captures the dynamics of the whole system, given in \eqs{eq:four_parameters} and \eq{eq:master_eq}.

In this article, we focus on the deceptively simple example of a homogeneous gas of two-level atoms and obtain the time-evolution of an initially fully inverted system. By calculating the radiated intensity and the off-diagonal coherence of the density matrix, we find a superradiant outburst at early times and subradiant emission immediately afterwards. The non-zero coherence shows that these are indeed cooperative effects. At late times, radiation trapping emerges \cite{holstein1947}, a regime characterized by a slowdown of emission in the absence of atomic coherences. Additionally, we demonstrate that observables such as the cooperative decay rate and the emission linewidth scale with the optical depth of the system. Finally, we investigate the slow decay in the subradiant regime, as well as the phase change of the radiated field during the cooperative decay. 

\section{Method}

The treatment in this section is a generalization and expansion of the method presented in Ref.~\cite{aamo2012}, with the comprehensive formalism  that is applicable to the presence of an external driving field, as well as an arbitrary geometry and particle distribution. It is worth noting that the system does not need to be small compared to the transition wavelength. Our treatment is valid as long as the time scale of cooperative decay $\tau_{coop}$ is much longer than the propagation time of the radiation through the media $l/c$, where $l$ is the size of the sample, and $c$ is the speed of light \footnote[1]{See Ref.~\cite{guerin2016}, for example, a typical experiment to observe cooperative effects. The size of the atomic cloud is $l\approx 1mm$. The propagation time of radiation is $l/c\approx 3\times 10^{-3}ns$. On the other hand, the shortest possible timescale of cooperative decay $\tau_{coop}\approx (\eta\g)^{-1}\approx 26ns/\eta$, where $\eta$ is the optical depth and $\g$ is the free-space natural linewidth.}.

\subsection{Model}

We consider an ensemble of $N$ identical two-level atoms that interact via a quantized electromagnetic field. We distinguish two probe atoms from the system, which are labeled as ``1'' and ``2''. With the dipole approximation, the full Hamiltonian of the system is given by

\be\label{eq:full_H}
	H=H_0+V
\ee
where
\beqq
	H_0&\!=\!&\sum_{i=1}^N \hbar\w_0 \s^\dagger_i \s_i\!+\!\sum_{\vec{k},\lambda}\hbar \w_l(\vec{k}) a^\dagger_{\vec{k},\lambda} a_{\vec{k},\lambda}\!-\!\!\sum_{i\ne1,2}\vec p_i\cdot (\vec {\cal E}_i+\vec E_i)\\
	V &\!=\!& -\!\sum_{i=1,2} \vec p_i\cdot (\vec {\cal E}_i+\vec E_i)
\eeqq
The first two terms in $H_0$ are the free Hamiltonians for the atoms and quantized field, respectively, with $\s^\dagger_i=\ket{e_i}\!\!\bra{g_i}$ and $\s_i=\ket{g_i}\!\!\bra{e_i}$ being the raising and lowering operators of the $i$-th atom, $a^\dagger$ and $a$ being the creation and annihilation operators of the photons corresponding to wavevector $\vec{k}$ and polarization $\lambda$. Here, $\vec E_i$ is the quantized field operator at the position of the $i$-th atom, while $\vec {\cal E}_i$ is the external classical driving field. $\vec p_i=\vec{\epsilon}_i \wp(\s_i+\s_i^\dagger)$ is the dipole operator of the $i$-th atom with real dipole matrix element $\wp$ and the unit polarization vector $\vec{\epsilon}_i$. Note that the spatial dependencies of these quantities are in the index $i$. The interaction term $V$ only includes the interaction between the probe atoms and the field, and all the atomic dipole operators couple to the radiation field at their own location.

Using the Keldysh formalism (see Ref.\cite{keldysh} and Appendix \ref{app:keldysh}), the dynamics of the full system are described by a time-evolution operator defined on the Schwinger-Keldysh contour $\cal C$.
\be \label{eq:SC}
	S_{\cal C} \;\equiv\; {\cal T}_{\cal C} e^{-\frac{i}{\hbar}\int\limits_{\cal C} d\check\tau \,V^I(\check\tau)}
\ee
where $V^I(\check\tau)$ is the interaction term $V$ in the interaction picture and the check mark on the time variable denotes a time on the Keldysh contour. Note that $V^I(\check\tau)$ contains the $\vec{E}_i$-term and therefore includes the effect of interactions with all the atoms. Each physical time $\tau$ corresponds to two times on the contour, $\check\tau=\tau_+$ on the upper branch and $\check\tau=\tau_-$ on the lower branch. $\TC$ is the time-ordering operator on the Keldysh contour defined as follows: it is the ordinary time-ordering operator ($\T$) on the upper branch, and the inverse time-ordering operator ($\T^{-1}$) on the lower branch. Additionally, terms that have time arguments on the lower branch are always ordered to the left, while terms on the upper branch are ordered to the right.

The state of the full system is described by a density matrix that includes the degrees of freedom of the $N$ atoms and the quantized field. Rewriting $\rho_{\alpha\beta}={\rm Tr}(\rho \ket{\beta}\!\!\bra{\alpha})$ in the interaction picture and defining the initial state $\rho_0\equiv\rho(t\!\to\!-\infty)$, the matrix elements can be written as (see Appendix \ref{app:keldysh})
\be \label{eq:rho}
	\rho_{\alpha\beta}(t) = {\rm Tr} \pare{\rho_0 \TC \SC P_{\beta\alpha}^I(t)} \equiv \sbrac{\TC \SC P_{\beta\alpha}^I(t)}_0
\ee
where $P^I_{\beta\alpha}(t)=e^{iH_0 (t-t_0)/\hbar}\ket{\beta}\!\!\bra{\alpha} e^{-iH_0 (t-t_0)/\hbar}$ is the projection operator $\ket{\beta}\!\!\bra{\alpha}$ written in the interaction picture.

\subsection{Master Equation of an Effective Two-atom System}

\subsubsection{Tracing out the Environment}

So far the model has been built upon an $N$-atom ensemble and the dimension of the density matrix in \eq{eq:rho} is, in principle, $2^{N}\times$ the number of field modes. The exponential increase of its dimensionality make it unfeasible for a large particle number. We hereby follow the idea of a mean-field approach by looking into two probe atoms, and treating the rest of $(N-2)$ atoms and the field as the environment (denoted by the subscript $E$).

We define the effective time-evolution operator for the two-atom system by taking the average with respect to the environment
\be \label{eq:SC1}
	\Se \equiv \sbrac{\SC}_E = \sbrac{\TC {\rm exp}\left\{ -\frac{i}{\hbar}\int_{\cal C} d\check\tau V^I(\check\tau)\right\}}_E
\ee
where $\sbrac{...}_E={\rm Tr}_E(...\rho_E)$ denotes an average over the environmental degrees of freedom.

Using a generalized cumulant expansion method for the operators (see Appendix \ref{app:cumulant}), \eq{eq:SC1} can be further expressed as
\beq 
	\Se &=& {\rm exp} \bigg\{\sum_{n=1}^\infty \frac{1}{n!}\pare{-\frac{i}{\hbar}}^n \int...\int_{\cal C}d\check\tau_1 ...d\check\tau_n\nonumber\\
	&&\times \dbrac{\TC V^I(\check\tau_1)...V^I(\check\tau_n)}_n\bigg\}
	\label{eq:SC2}
\eeq
where $\dbrac{\TC V^I(\check\tau_1)...V^I(\check\tau_n)}_n$ is the $n$-th order cumulant. The second-order cumulant, for example, is defined as $\dbrac{AB}_2\!\equiv\! \sbrac{AB}\!-\!\sbrac{A}\!\sbrac{B}$. Note that each of the cumulants is assigned with a time-ordering operator, such that the $V^I$ terms are ordered according to the branches that $\check{\tau}_1...\check{\tau}_n$ sit on. We further assume a Gaussian form for the radiation field, that is, only cumulants up to the second-order are kept \footnote[2]{In a series expansion, we cut off after the second-order cumulants of the field operators. Since the second-order cumulants will be calculated self-consistently by an average of the full system, they indeed include the higher-order contribution that is constructed by the two-body interaction. In the simplest case where there is no true many-body interactions, the treatment here is exact}. The resulting two-atom density matrix is a $4\times 4$ matrix
\beq
\rho_{\alpha\beta}(t)&=& \sbrac{\TC \Se P^I_{\beta\alpha}(t)}_0 \label{eq:end1}
\eeq
where each of the subscripts $\alpha$ and $\beta$ represent a bare two-atom eigenstates $\alpha,\beta\in$ \{gg,ge,eg,ee\}.

\subsubsection{Rotating Wave Approximation}

In this section, we express the cumulants in terms of atomic dipole operators and quantized field operators. Note that all time-dependent operators in this section are in the interaction picture and that we drop the superscript to simplify the notation.
We first separate the dipole and the fields into slow varying positive and negative frequency components 

\beqq
p_{i\mu}(\tau) &=& p_{i\mu}^+(\tau) e^{-i\omega_0\tau} + p_{i\mu}^-(\tau) e^{i\omega_0\tau}\\
E_{i\mu}(\tau) &=& E_{i\mu}^+(\tau)e^{-i\omega_l\tau} + E_{i\mu}^-(\tau)e^{i\omega_l\tau}\\
{\cal E}_{i\mu}(\tau) &=& {\cal E}_{i\mu}^+(\tau)e^{-i\omega_c\tau} + {\cal E}_{i\mu}^-(\tau)e^{i\omega_c\tau}
\eeqq
where $i=1,2$ is the index for the probe atom and $\mu=x,y,z$ are the spatial components. Note that $p_{i\mu}$ and $E_{i\mu}$ are operators, while the classical driving field ${\cal E}_{i\mu}$ is a complex number. The negative (positive) frequency component of the dipole operator is associated with the atomic raising (lowering) operator
\beq
	p_{i\mu}^+(t) &=& \wp_\mu \s_{i\mu}(t)\\
	p_{i\mu}^-(t) &=& \wp_\mu \s_{i\mu}^\dagger(t)
\eeq
where $\wp_\mu$ is the dipole matrix element for polarization $\mu$ and $\s_{i\mu}=\ket{g}_{i\mu}\!\!\bra{e}$.

Using the rotating wave approximation to drop the fast-rotating terms, we obtain the first-order cumulant as
\beqq
\dbrac{V(\check\tau)} &=& \sbrac{V(\check\tau)}\\
&=& -\sum_{i,\mu}\left(p^+_{i\mu}(\check\tau){\cal E}^-_{L,i\mu}(\check\tau) + p^-_{i\mu}(\check\tau){\cal E}^+_{L,i\mu}(\check\tau) \right)
\eeqq
where we have defined the Lorentz-Lorenz local field \cite{lorentz,lorenz}
\be
{\cal E}_{L,i\mu}^\pm \;\equiv\; {\cal E}_{i\mu}^\pm + \langle E_{i\mu}^\pm \rangle
\ee
The second-order term is calculated as follows
\beq
\lefteqn{\dbrac{\TC V(\check\tau_1)V(\check\tau_2)}\;=}\\
&=& \sum_{i\mu,j\nu}\dbrac{\TC\; p_{i\mu}(\check\tau_1)E_{i\mu}(\check\tau_1) p_{j\nu}(\check\tau_2)E_{j\nu}(\check\tau_2)}\nonumber\\
&=& \sum_{i\mu,j\nu} \{p^+_{i\mu}(\check\tau_1)\dbrac{{\cal T}_{\cal C}E^-_{i\mu}(\check\tau_1)E^+_{j\nu}(\check\tau_2)}p^-_{j\nu}(\check\tau_2)e^{-i\omega(\check\tau_1-\check\tau_2)}\nonumber\\
&& + p^-_{i\mu}(\check\tau_1)\dbrac{{\cal T}_{\cal C}E^+_{i\mu}(\check\tau_1)E^-_{j\nu}(\check\tau_2)}p^+_{j\nu}(\check\tau_2)e^{i\omega(\check\tau_1-\check\tau_2)} \}\nonumber
\eeq
where $\omega = \omega_0-\omega_l$ is the detuning between atomic frequency and light frequency. In the first step, all terms with classical field $\cal E$ vanish in the cumulant. In the second step, the dipole operators are moved outside the cumulants, since the average does not involve the probe atoms. Note that we explicitly write the time-ordering operator ${\cal T}_{\cal C}$ within the cumulants.

We further define the following quantities
\begin{subequations}\label{eq:greens_function}
\begin{align}
D_{i\mu,j\nu}^{\pm\pm} &\equiv \dbrac{{\cal T}_{\cal C} E^-_{i\mu}(\check\tau_{1\pm})E^+_{j\nu}(\check\tau_{2\pm})}\\
C_{i\mu,j\nu}^{\pm\pm} &\equiv \dbrac{{\cal T}_{\cal C} E^+_{i\mu}(\check\tau_{1\pm})E^-_{j\nu}(\check\tau_{2\pm})}
\end{align}
\end{subequations}
where the double superscript of $D$ and $C$ stands for the branches of $\check\tau_1$ and $\check\tau_2$. $D$ stands for terms like $\dbrac{{\cal T}_{\cal C}E^-E^+}$, while $C$ is for $\dbrac{{\cal T}_{\cal C}E^+E^-}$. The exact form of ${\cal T}_{\cal C}$ depends on the branches of $\check\tau_1$ and $\check\tau_2$. For example, $D_{i\mu,j\nu}^{--} = \dbrac{{\cal T}^{-1} E^-_{i\mu}(\check\tau_{1-})E^+_{j\nu}(\check\tau_{2-})}$ because both of the times are in the lower branch.

\subsubsection{Derivation of the Master Equation}

Taking the time derivative on both sides of Eq.~(\ref{eq:end1}) results in the master equation
\be \label{eq:maeq}
\dot \rho_{\alpha\beta}(t) = \sbrac{\TC \dot\Se P^I_{\beta\alpha}(t)} + \sbrac{\TC \Se \dot P^I_{\beta\alpha}(t)}
\ee
The second term of \eq{eq:maeq} gives rise to the dynamics corresponding to the free Hamiltonian $H_0$
\beqq
\sbrac{\TC \Se \dot P^I_{\beta\alpha}(t)}&=& -\frac{i}{\hbar}\sbrac{\TC \Se \squ{P^I_{\beta\alpha}(t),H_0}}\\
&=& \ih \bra{\alpha}\squ{H_0,\rho(t)}\ket{\beta}
\eeqq
The first term of \eq{eq:maeq} includes all the non-trivial effects arising from second-order correlations. Defining the term
\be
P \equiv \sum_{n=1}^\infty \frac{1}{n!}\pare{-\frac{i}{\hbar}}^n \int ... \int_{\cal C}d\check\tau_1 ...d\check\tau_n \dbrac{\TC V^I(\check\tau_1)...V^I(\check\tau_n)}_E
\ee
one can formally write
\begin{widetext}
\be\label{eq:Sceff}
\dot \Se = \frac{d}{dt}\TC e^P = \TC \pare{\frac{d}{dt}P}e^P= \TC \frac{d}{dt}\pare{\ih \int_{\cal C}d\check\tau_1 \sbrac{V^I(\check\tau_1)} - \frac{1}{2\hbar^2} \iint_{\cal C}d\check\tau_1 d\check\tau_2 \dbrac{\TC V^I(\check\tau_1)V^I(\check\tau_2)}+...}e^P.
\ee
The $t-$dependence in $P$ goes into the integration limits: $\int_{\cal C}d\check\tau=\int_{-\infty}^td\tau_++\int^{-\infty}_td\tau_-$. Note that the Keldysh contour can be extended to $+\infty$ with the present time $t$ assigned to the lower branch, such that the integration becomes $\int_{\cal C}d\check\tau=\int_{-\infty}^{+\infty}d\tau_++\int_{+\infty}^{t}d\tau_-+\int^{-\infty}_td\tau_-$. Under the Gaussian-field assumption, we only keep the terms with first and second order cumulants. Then, the first term of \eq{eq:maeq} is simply given by
\begin{eqnarray*}
\sbrac{\TC \dot\Se P^I_{\beta\alpha}(t)}=&& \frac{i}{\hbar} \sum_{i,\mu}\wp_\mu \bra{\alpha}\squ{\sigma_{i\mu}{\cal E}^-_{L,i\mu} + \sigma_{i\mu}^\dagger{\cal E}^+_{L,i\mu},\rho(t)}\ket{\beta}\\
&-&\sum_{i\mu,j\nu}\frac{\wp_\mu \wp_\nu}{\hbar^2} \int^{+\infty}_0 d\tau \bra{\alpha}\bigg\{\sigma_{i\mu}\sigma_{j\nu}^{\dagger}\rho(t)f^{(-)}_{i\mu j\nu}(\tau)
+ \sigma_{j\nu}^{\dagger}\sigma_{i\mu}\rho(t)g^{(+)}_{i\mu j\nu}(\tau)\\
&+& \rho(t)\sigma_{j\nu}^\dagger\sigma_{i\mu}g^{(-)}_{i\mu j\nu}(\tau)
+ \rho(t)\sigma_{i\mu}\sigma_{j\nu}^\dagger f^{(+)}_{i\mu j\nu}(\tau) - \sigma_{i\mu}\rho(t)\sigma_{j\nu}^\dagger g^{(-)}_{i\mu j\nu}(\tau)
- \sigma_{j\nu}^\dagger\rho(t)\sigma_{i\mu}f^{(+)}_{i\mu j\nu}(\tau)\\
&-& \sigma_{i\mu}\rho(t)\sigma_{j\nu}^\dagger g^{(+)}_{i\mu j\nu}(\tau)
- \sigma_{j\nu}^\dagger\rho(t)\sigma_{i\mu}f^{(-)}_{i\mu j\nu}(\tau) \bigg\}\ket{\beta}
\end{eqnarray*}
\end{widetext}
The two-time correlation functions are defined as
\begin{subequations}\label{eq:cumulants}
\begin{align}
f^{(-)}_{i\mu j\nu}(\tau) &\equiv e^{-i\omega\tau} \dbrac{E_{i\mu}^-(t)E_{j\nu}^+(t-\tau)} \label{eq:f_minus}\\
f^{(+)}_{i\mu j\nu}(\tau) &\equiv e^{i\omega\tau} \dbrac{E_{i\mu}^-(t)E_{j\nu}^+(t+\tau)}\label{eq:f_plus}\\
g^{(-)}_{i\mu j\nu}(\tau) &\equiv e^{-i\omega\tau} \dbrac{E_{j\nu}^+(t-\tau)E_{i\mu}^-(t)} \\
g^{(+)}_{i\mu j\nu}(\tau) &\equiv e^{i\omega\tau} \dbrac{E_{j\nu}^+(t+\tau)E_{i\mu}^-(t)} \label{eq:g_plus}
\end{align}
\end{subequations}
where $\w= \omega_0-\omega_l$ can be seen as a Fourier variable. A detailed derivation of the result above can be found in Appendix \ref{AppC}. We additionally define the following quantities

\begin{subequations}\label{eq:four_parameters}
\begin{align}
&\Gamma_{i\mu,j\nu}(\omega,t) =\frac{\wp_\mu \wp_\nu}{\hbar^2} \int\limits^{+\infty}_0 d\tau (f^{(+)}_{i\mu j\nu}(\tau) + f^{(-)}_{i\mu j\nu}(\tau) )\label{eq:GG1}\\
&\gamma_{i\mu,j\nu}(\omega,t) + \Gamma_{i\mu,j\nu}(\omega,t) =\nonumber\\
&\quad\quad\quad\quad \frac{\wp_\mu \wp_\nu}{\hbar^2} \int\limits^{+\infty}_0 d\tau (g^{(+)}_{i\mu j\nu}(\tau) + g^{(-)}_{i\mu j\nu}(\tau) )\\
&\Delta_{i\mu,j\nu}(\omega,t) = -\frac{i}{\hbar^2} \frac{\wp_\mu \wp_\nu}{2} \int\limits^{+\infty}_0 d\tau (f^{(+)}_{i\mu j\nu}(\tau) - f^{(-)}_{i\mu j\nu}(\tau))\\
&\delta_{i\mu,j\nu}(\omega,t) + \Delta_{i\mu,j\nu}(\omega,t) =\nonumber\\
&\quad\quad\quad\quad -\frac{i}{\hbar^2} \frac{\wp_\mu \wp_\nu}{2} \int\limits^{+\infty}_0 d\tau (g^{(+)}_{i\mu j\nu}(\tau) - g^{(-)}_{i\mu j\nu}(\tau))
\end{align}
\end{subequations}
Here, the $\Gamma$ and $\gamma$ terms describe broadening (decay rates) and will consequently show up in the Lindblad part of the master equation. From the mathematical form, one can see that all $\Gamma$ terms are proportional to the light intensity, and thus constitute terms akin to stimulated emission/absorption and power broadening.  The $\gamma$ terms, on the other hand, describe spontaneous decay. In vacuum, they follow the Wigner-Weisskopf form of free-space natural linewidth \cite{ms1999} for $i\mu=j\nu$, and are $0$ for $i\mu\ne j\nu$. The terms $\Delta$ and $\delta$ appear in the Hamiltonian part of the master equation, and describe collectively induced shifts and Lamb shifts, respectively.

Note that $\G_{i\mu j\nu}$ and $\Delta_{i\mu j\nu}$ satisfy the relation
\be
\frac{\wp_\mu \wp_\nu}{\hbar^2} \int^{+\infty}_0 d\tau f^{(\pm)}_{i\mu j\nu}(\tau) = \frac{\Gamma_{i\mu,j\nu}(\omega,t)}{2} \pm i \Delta_{i\mu,j\nu}(\omega,t)\label{eq:analytic}
\ee
As $\G_{i\mu j\nu}$ and $\Delta_{i\mu j\nu}$ are the real and imaginary parts of this analytic function, they are connected via the Kramers-Kronig relation, such that $\Delta_{i\mu j\nu}$ can be alternatively computed as
\be\label{eq:light_shift}
\Delta_{i\mu,j\nu}(\w,t) = -\frac{1}{2\pi}{\cal P}\int_{-\infty}^{+\infty}\frac{\G_{i\mu,j\nu}(\w',t)}{\w'-\w}d\w'
\ee
where $\cal P$ denotes a principle value integral. An analogous relation exists for the pair $\g_{i\mu j\nu}$ and $\delta_{i\mu j\nu}$.

Then, the master equation for the two-atom system in the rotating frame finally reads

\begin{eqnarray}\label{eq:master_eq}
\dot \rho &=& \ih \squ{\tilde{H}_0,\rho}+\frac{i}{\hbar} \sum_{i,\mu}\wp_\mu \squ{\sigma_{i\mu}{\cal E}^-_{L,i\mu} + \sigma_{i\mu}^\dagger{\cal E}^+_{L,i\mu},\rho}\nonumber\\
&&+i\sum_{i,\mu\nu}\Delta_{i\mu,i\nu}\squ{\squ{\s_{i\mu},\s^\dagger_{i\nu}},\rho}-i\sum_{i\mu,j\nu}\delta_{i\mu,j\nu}\squ{\s^\dagger_{j\nu}\s_{i\mu},\rho}\nonumber\\
&& -\sum_{i\mu,j\nu} \frac{\Gamma_{i\mu,j\nu}}{2} (\sigma_{i\mu}\sigma_{j\nu}^\dagger\rho - 2\sigma_{j\nu}^\dagger\rho\sigma_{i\mu} + \rho\sigma_{i\mu}\sigma_{j\nu}^\dagger) \nonumber\\
&&-\sum_{i\mu,j\nu} \frac{\Gamma_{i\mu,j\nu} + \gamma_{i\mu,j\nu}}{2} (\sigma_{j\nu}^\dagger\sigma_{i\mu}\rho - 2\sigma_{i\mu}\rho\sigma_{j\nu}^\dagger +\rho\sigma_{j\nu}^\dagger\sigma_{i\mu})\nonumber\\
\end{eqnarray}
\noindent where $\tilde{H}_0=\hbar(\w_0-\w_c)\sum_{i}\s^\dagger_{i}\s_{i}=\hbar\Delta_0\sum_{i}\s^\dagger_{i}\s_{i}$ is the two-atom free Hamiltonian in the rotating frame. This master equation together with \eq{eq:four_parameters} are the first major results of this article.

\subsection{Closed Form Expressions for $\G$}

While we have formally obtained the two-atom master equation in \eq{eq:master_eq}, the specific expressions of the cumulants in \eqs{eq:cumulants} and \eqs{eq:four_parameters} are still unknown. In this section, we show how to calculate the cumulants, and further express the cooperative decay rate $\G_{i\mu j\nu}$ and the light shift $\Delta_{i\mu j\nu}$.

This amounts to a closed form for solving $\G_{i\mu j\nu}$ and $\Delta_{i\mu j\nu}$, as shown in \fig{fig:closed_form}. The procedure is briefly summarized as follows: (a) We rewrite the master equation in \eq{eq:master_eq} as the Maxwell-Bloch equations in \eq{eq:nine_eqs}. (b) We employ a Dyson equation formalism \cite{dyson1949} to relate the cumulants of the field operators in \eqs{eq:cumulants} to their source, the atomic polarization functions, which depend on the cumulants of the atomic transition operators $\sdbrac{\s'_i(t_1)\s''_j(t_2)}$, where $\s',\s''$ are either raising ($\s^\dagger$) or lowering operator ($\s$) of the atomic transition. (c) The cooperative decay rate $\G_{i\mu j\nu}$ in \eq{eq:GG1} is then expressed in terms of $\sdbrac{\s'_i(t_1)\s''_j(t_2)}$. (d) The cumulants $\sdbrac{\s'_i(t_1)\s''_j(t_2)}$ are calculated using the quantum regression theorem, and expressed in terms of the six atomic variables in \eqs{eq:atomic_vars}. (e) Finally, $\G_{i\mu j\nu}$ and $\Delta_{i\mu j\nu}$ are written in terms of the six atomic variables, allowing the time-evolution in \eq{eq:nine_eqs} to be solved self-consistently.

\subsubsection{Simplification of the Master Equation}

We first rewrite the two-atom master equation \eq{eq:master_eq} in terms of the matrix elements, by introducing the following notation

\be
\rho_{\alpha\beta,\gamma\delta} \equiv \bra{\alpha\gamma}\rho\ket{\beta\delta}
\label{eq:matrix_element}
\ee
where $\ket{\beta\delta} \equiv \ket{\beta}_1 \ket{\delta}_2$ is the product of the single-atom states, namely $\ket{\beta\delta} \in \{\ket{gg},\ket{eg},\ket{ge},\ket{ee}\}$. We define the following atomic variables as combinations of the matrix elements
\begin{subequations}\label{eq:atomic_vars}
\begin{align}
a &\equiv (\sbrac{\s_{ee1}}+\sbrac{\s_{ee2}})/2\nonumber\\
&=\rho_{ee,ee}+(\rho_{ee,gg}+\rho_{gg,ee})/2\\
n &\equiv \sbrac{\s_{z1}\s_{z2}}\nonumber\\
&=\rho_{ee,ee}-\rho_{ee,gg}-\rho_{gg,ee}+\rho_{gg,gg}\\
x &\equiv (\langle\s_1 \s_2^\dagger\rangle+\langle\s
_1^\dagger \s_2\rangle)/2\nonumber\\
&=(\rho_{eg,ge}+\rho_{ge,eg})/2\\
\rho_{eg} &\equiv(\sbrac{\s_1}+\sbrac{\s_2})/2\nonumber\\
&=(\rho_{eg,ee}\!+\!\rho_{eg,gg}\!+\!\rho_{ee,eg}\!+\!\rho_{gg,eg})/2=\rho_{ge}^*\\
m_{eg} &\equiv (\sbrac{\s_1 \s_{z2}}+\sbrac{\s_{z1}\s_2})/2\nonumber\\
&=(\rho_{eg,ee}\!-\!\rho_{eg,gg}\!+\!\rho_{ee,eg}\!-\!\rho_{gg,eg})/2=m_{ge}^*\\
\rho_{eg,eg}&\equiv\sbrac{\s_1 \s_2}=\rho_{ge,ge}^*
\end{align}
\end{subequations}
\noindent where $\s_{eei}\equiv \ket{e}_i\!\!\bra{e}$ and $\s_{zi}$ is the Pauli Z operator of the $i-$th atom.
$a$ is the average excited state population, and $n$ is the effective two-atom inversion. Further assuming a permutational symmetry for atom ``1" and atom ``2", there are 6 independent variables left for a Hermitian matrix with trace one. We neglect the retardation effects of electromagnetic wave propagation \cite{aamo2012}, and therefore, the spatial dependence of all variables can be omitted. We consider a single polarization component of the field, so that the subscript $\mu,\nu$ are dropped. Then, the master equation can be written as the following set of Maxwell-Bloch equations
\begin{widetext}
\begin{subequations}\label{eq:nine_eqs}
\begin{align}
\dot{a} &= \G-(\g+2\G)a-\bar{\g}x-i\W(\rho_{eg}-\rho_{ge})\label{eq:a}\\
\dot{n} &= 2\g-2(\g+2\G)n-4\g a+4(\bar{\g}+2\bar{\G})x-4i\W(m_{eg}-m_{ge})\\
\dot{x} &= -(\g+2\G )x + \frac{\bar{\g}+2\bar{\G}}{2}n + \bar{\g}a - \frac{\bar{\g}}{2} + i\W(m_{eg}-m_{ge})\\
\dot{\rho}_{eg} &= (\dot{\rho}_{ge})^*=-\bigg(\frac{\g+2\G}{2}+i(\delta+2\Delta+\Delta_0)\bigg)\rho_{eg} + \frac{\bar{\g}+2i\bar{\delta}}{2} m_{eg}-i\W(2a-1)\label{eq:reg}\\
\dot{m}_{eg} &= (\dot{m}_{ge})^*=-\bigg(3\frac{\g+2\G}{2}+\bar{\g}+2\bar{\G}+i(\delta+2\Delta+\Delta_0)\bigg)m_{eg} -(\g+\frac{\bar{\g}}{2}-i\bar{\delta})\rho_{eg}-i\W(n-2x+2\rho_{egeg})\\
\dot{\rho}_{egeg} &= (\dot{\rho}_{gege})^*=-((\g+2\G )+2i(\delta+2\Delta+\Delta_0))\rho_{egeg}-2i\W m_{eg}
\end{align}
\end{subequations}
\end{widetext}
where $\G=(\G_{11}+\G_{22})/2$ is the induced pump and decay rate from a single atom, and $\bar{\G}=(\G_{12}+\G_{21})/2$ is the inter-atom contribution. $\g=(\g_{11}+\g_{22})/2$ and $\bar{\g}=(\g_{12}+\g_{21})/2$ are the spontaneous decay rates. Similarly, $\Delta=(\Delta_{11}+\Delta_{22})/2$ is the induced light shift, while $\delta=(\delta_{11}+\delta_{22})/2$ and  $\bar{\delta}=(\delta_{12}+\delta_{21})/2$ are the spontaneous light (Lamb) shifts. Since the collectively modified spontaneous decay rate $\g$ is very close to the free-space spontaneous emission rate $\g_f$ \cite{louisell}, and the collective Lamb shift $\delta$ is very close to the free-space Lamb shift $\delta_l$, we hereby replace $\g$ by $\g_f$, and $\delta$ by $\delta_l$, and set $\bar{\g}=\bar{\delta}=0$. Additionally, $\Omega=|{\cal E}_L^{\pm}|\wp/\hbar$ and $\Delta_0=\w_0-\w_c$ denote the Rabi frequency and the detuning of the external driving field, respectively. The physical interpretation of the parameters are summarized in \tab{tab:parameters}.

\begin{table}
\begin{tabular}{|m{14em}|c|c|}
\hline
& single-atom & inter-atom \\
\hline
\\[-1em]
induced pump and decay rate & $\G$ & $\bar{\G}$ \\
\hline
\\[-1em]
spontaneous decay rate & $\g\approx\g_f$ & $\bar{\g}\approx 0$ \\
\hline
\\[-1em]
induced light shift & $\Delta$ & $\bar{\Delta}$ \\
\hline
\\[-1em]
spontaneous light shift & $\delta\approx\delta_l$ & $\bar{\delta}\approx 0$ \\
\hline
\end{tabular}
\caption{\label{tab:parameters}Summarising parameters in \eqs{eq:nine_eqs}. $\g_f$ is the free-space spontaneous emission rate, and $\delta_l$ is the free-space Lamb shift.}
\end{table}

\subsubsection{Cooperative Decay Rate $\G$ via Dyson Equation Formalism}

From \eq{eq:f_minus}, \eq{eq:f_plus} and \eq{eq:GG1}, the cooperative decay rate $\G$ depends on the cumulants of the form
\bee
D_{1\mu,2\nu}^{-+} = \dbrac{{\cal T}_{\cal C} E^-_{1\mu}(\check\tau_{1-})E^+_{2\nu}(\check\tau_{2+})}
\eee
Using the Dyson equation formalism \cite{fetter}, they can be formally expressed as
\beq
D_{1\mu,2\nu}(\check{1},\check{2})&=&D_{1\mu,2\nu}^{(0)}(\check{1},\check{2})-\sum_{\alpha,\beta}\int\!\!\int d\check{1}'d\check{2}' D_{1\mu,1'\alpha}^{(0)}(\check{1},\check{1}')\nonumber\\
&&\times\Pi_{1'\alpha,2'\beta}(\check{1}',\check{2}')D_{2'\beta,2\nu}(\check{2}',\check{2})\label{eq:dyson}
\eeq
where $\check{1}\equiv (\vec{r}_1,\check{\tau}_1)$, $\check{2}\equiv (\vec{r}_2,\check{\tau}_2)$, and $D_{1\mu,2\nu}^{(0)}(\check{1},\check{2})\!=\!\dbrac{\TC E_{1\mu}^{(0)-}(\check{1})E_{2\nu}^{(0)+}(\check{2})}$ is the free space Green's function. The integration goes over the whole space and the complete Keldysh contour. The source function $\Pi$ comes from the following ansatz which corresponds to a self-consistent Hartree approximation:
\beq\label{eq:ansatz}
\Pi_{1\alpha,2\beta}(\check{1},\check{2})&=&\frac{\wp_\alpha \wp_\beta}{\hbar^2}\sum_{i,j}\dbrac{\TC \s^\dagger_{i\alpha}(\check{\tau}_1)\s_{j\beta}(\check{\tau}_2)}\nonumber\\
&&\times\delta(\vec{r}_1-\vec{r}_i)\delta(\vec{r}_2-\vec{r}_j)
\eeq
Therefore, the cooperative decay rate $\G_{i\mu j\nu}$ is related to the cumulants of atomic transition operators. The single-atom decay rate $\G$ and the inter-atom decay rate $\bar{\G}$ in Fourier space are readily obtained as the following, with the detailed derivation available in Ref.~\cite{aamo2012,ms1999}

\begin{widetext}
\begin{subequations}\label{eq:G_Gb}
\begin{align}
\G(\w,t) =& \frac{\wp^2}{\hbar^2}\int_V d^3x |\tilde{D}^{ret}(\vec{r},\vec{x},\w,t)|^2 P^{(1)s}(\vec{x},\w,t)\nonumber\\
&\quad\quad+ \frac{\wp^2}{\hbar^2}\iint_V d^3x_1 d^3x_2 \tilde{D}^{ret}(\vec{r},\vec{x}_1,\w,t)\tilde{D}^{*ret}(\vec{r},\vec{x}_2,\w,t)P^{(2)s}(\vec{x}_1,\vec{x}_2,\w,t)\nonumber\\
\equiv &A_1+B\\
\bar{\G}(\w,t) =& \frac{\wp^2}{\hbar^2}\int_V d^3x \tilde{D}^{ret}(\vec{r}_1,\vec{x},\w,t)\tilde{D}^{*ret}(\vec{r}_2,\vec{x},\w,t)P^{(1)s}(\vec{x},\w,t)\nonumber\\
&\quad\quad+ \frac{\wp^2}{\hbar^2}\iint_V d^3x_1 d^3x_2 \tilde{D}^{ret}(\vec{r}_1,\vec{x}_1,\w,t)\tilde{D}^{*ret}(\vec{r}_2,\vec{x}_2,\w,t)P^{(2)s}(\vec{x}_1,\vec{x}_2,\w,t)\nonumber\\
\equiv& A_2 +B'
\end{align}
\end{subequations}
\end{widetext}
where the retarded Green's function in the medium takes the form (see Appendix \ref{App:retarded_greens_function})
\begin{eqnarray}
\tilde{D}^{ret}(\vec{x}_1,\vec{x}_2,\w,t) &=& -\frac{i\hbar\w_0^2}{6\pi\epsilon_0 c^2}\frac{e^{-iq_0(\w)r}}{r}  \label{eq:q0main} \\
{\rm with}\quad q_0(\w) &=&\frac{\w_0}{c}\bigg(1+i\frac{\hbar}{3\epsilon_0 }P^{(1)ret}(\vec{r},\w,t)\bigg)\nonumber
\end{eqnarray}
where $r=|\vec{x}_1-\vec{x}_2|$, and $\w_0$ is the atomic transition frequency. To derive \eq{eq:q0main}, we assume that the source function $P^{(1)ret}$ is small. This approximation is justified in Appendix \ref{App:taylor_expansion} for a non-driven system.

The source function $\Pi$ in \eq{eq:ansatz} can be written in a continuous approximation, and in \eqs{eq:G_Gb} and \eq{eq:q0main} we should use the Fourier transform (with respect to $\tau\!=\!t_1\!-\!t_2$) of the following expressions:
\begin{subequations}\label{eq:source}
\begin{align}
&P^{(1,2)s}(r_i,r_j;t_1,t_2) = \frac{\wp^2}{\hbar^2} N^{1(2)} \dbrac{\s^\dagger_i(t_1)\s_j(t_2)}\label{eq:source1}\\
&P^{(1,2)ret}(r_i,r_j;t_1,t_2) =\nonumber\\
&\quad\quad\frac{\wp^2}{\hbar^2} N^{1(2)} \theta(t_1-t_2) \dbrac{\squ{\s^\dagger_i(t_1),\s_j(t_2)}}\label{eq:source2}
\end{align}
\end{subequations}
where the superscripts ``1'' and ``2'' stand for the one-atom ($i=j$) and the two-atom ($i \neq j$) source functions, respectively. Additionally, $N$ denotes the particle density of the sample, and $\theta(t)$ is the Heaviside step function.

\subsubsection{Cumulants of Atomic Transition Operators and the Source Functions}

In this section, we derive the source functions in \eqs{eq:source}, which will be written in terms of the six atomic variables in \eqs{eq:atomic_vars}.

By virtue of the quantum regression theorem \cite{meystre}, the equation of motion for the two-time correlation function is the same as that for the one-time expectation value. For example, directly from \eq{eq:reg} we obtain 
\beq
\frac{d}{d\tau}\sbrac{\s(t+\tau)}&=&-(\g/2+\G+i\tilde{\delta})\sbrac{\s(t+\tau)}\nonumber\\
&&-i\Omega(2\sbrac{\s_{ee}(t+\tau)}-1),
\eeq
where we have defined the total detuning $\tilde{\delta}=\delta+2\Delta+\Delta_0$, and denote the excited state projection operator $\ket{e}\!\!\bra{e}$ by $\s_{ee}$. Then, for an arbitrary time-dependent atomic operator $\pi(t)$, we have
\beq
\frac{d}{d\tau}\dbrac{\pi(t)\s(t+\tau)}&=&-(\g/2+\G+i\tilde{\delta})\dbrac{\pi(t)\s(t+\tau)}\nonumber\\
&&-2i\Omega\dbrac{\pi(t)\s_{ee}(t+\tau)}\label{eq:beforeLT}
\eeq
where we have used the definition of the second-order cumulant $\dbrac{AB}\!\equiv\! \sbrac{AB}\!-\!\sbrac{A}\!\sbrac{B}$. Note that the constant term vanishes because $\dbrac{\pi\cdot 1}=\sbrac{\pi\cdot 1}-\sbrac{\pi}\sbrac{1}=0$. Defining the function $f_{\pi\s}(\tau;t)\equiv\dbrac{\pi(t)\s(t+\tau)}$ and its Laplace transform with respect to $\tau$
\be
F_{\pi\s}(\lambda;t)\!\equiv\! LT\{f_{\pi\s}(\tau;t)\}(\lambda)\!=\!\!\int_0^\infty \!\!f_{\pi\s}(\tau;t) e^{-\lambda \tau}d\tau
\ee
We can rewrite Eq.(\ref{eq:beforeLT}) as
\beq
&&\lambda F_{\pi\s}(\lambda;t)-f_{\pi\s}(0;t)\nonumber\\
&=& -(\g/2+\G+i\tilde{\delta}) F_{\pi\s}(\lambda;t) - 2i\Omega F_{\pi e}(\lambda;t)\label{eq:LT1}
\eeq
where $f_{\pi\s}(0;t)\!\!=\!\!\dbrac{\pi(t)\s(t)}$ is the initial condition, and $F_{\pi e}(\lambda;t)$ is the Laplace transform of $f
_{\pi e}(\tau;t)\!\!\equiv\!\! \dbrac{\pi(t)\s_{ee}(t+\tau)}$. Similarly, we obtain the corresponding equations for $a$ and $\rho_{ge}$
\beq
&&\lambda F_{\pi e}(\lambda;t)-f_{\pi e}(0;t)\nonumber\\
&=& -(\g+2\G)F_{\pi e}(\lambda;t)\!-\! i\W\big[F_{\pi\s}(\lambda;t)\!-\! F_{\pi\s^\dagger}(\lambda;t)\big]
\eeq
and
\beq
&&\lambda F_{\pi\s^\dagger}(\lambda;t)-f_{\pi\s^\dagger}(0;t)\nonumber\\
&=& -(\g/2+\G-i\tilde{\delta}) F_{\pi\s^\dagger}(\lambda;t) + 2i\Omega F_{\pi e}(\lambda;t)\label{eq:LT3}
\eeq
Combining Eq.~(\ref{eq:LT1})-(\ref{eq:LT3}), we can solve for $F_{\pi\s}$, $F_{\pi\s^\dagger}$ and $F_{\pi e}$, where the operator $\pi$ can be replaced by $\s$ or $\s^\dagger$. To obtain the expressions in the Fourier space, it suffices to replace $\lambda\!\to\! i \omega$. The approach above allows us to obtain the Fourier transform of the following cumulants: $\sdbrac{\s_j(t)\s_i^\dagger(t+\tau)}$, $\sdbrac{\s_i^\dagger(t+\tau)\s_j(t)}$, $\sdbrac{\s_j^\dagger(t)\s_i(t+\tau)}$ and $\sdbrac{\s_i(t+\tau)\s_j^\dagger(t)}$. Substituting the Fourier transform of the cumulants into \eqs{eq:source}, we obtain the source functions
\begin{widetext}
\beq
P^{(1,2)s} &=& 2 \frac{\wp^2}{\hbar^2}N^{1(2)} Re\bigg[
\frac{2\W A_0 (i\G_f-\w-\tilde{\delta})+R_{ge0}(2\G_f+i\w)(\G_f+i\w+i\tilde{\delta})+2\W^2(R_{eg0}+R_{ge0})}
{(2\G_f+i\w)((\G_f+i\w)^2+\tilde{\delta}^2)+4\W^2(\G_f+i\w)}
\bigg]\\
P^{(1)ret} &=& \frac{\wp^2}{\hbar^2}N
\frac{(2a-1)((\G_f+i\tilde{\delta}+i\w)(2\G_f+i\w)+2\W^2)+2\W\rho_{eg}(-i\G_f+\tilde{\delta}+\w)}
{(\tilde{\delta}^2+(\G_f+i\w)^2)(2\G_f+i\w)+4\W^2(\G_f+i\w)}
\eeq
\end{widetext}
where $\G_f=\G+\g/2$. For $P^{(1)s}$, $A_0 = -a\rho_{eg}$, $R_{ge0}=a-\rho_{eg}\rho_{ge}$, $R_{eg0}=-\rho_{eg}^2$. For $P^{(2)s}$, $A_0 = \rho_{ee,eg}-a\rho_{eg}=\frac{\rho_{eg}+m_{eg}}{2}-a\rho_{eg}$, $R_{ge0}=\rho_{eg,ge}-\rho_{eg}\rho_{ge}$ and $R_{eg0}=\rho_{eg,eg}-\rho_{eg}^2$.

\subsubsection{The Explicit Form of $\G$ in Terms of Atomic Variables}

Approximately, the spatial integrations for $\Gamma$ and $\bar{\Gamma}$ in \eqs{eq:G_Gb} are done over a sphere centered at the location of the probe atoms, with diameter $l$ equaled to the size of the sample.
\bee
\int_V d^3x \to \int_0^{2\pi}d\phi\int_0^\pi \sin\theta d\theta \int_0^{l/2}r^2  dr
\eee

Using the Wigner-Weisskopf result for spontaneous decay, we do the substitution $\wp^2 \w^3_0/3\pi \hbar \epsilon_0 c^3 \!\to\!\g$. Along with the transition wavelength $\lambda=2\pi c/\w_0$, we define the {\it particle number within a cubed wavelength}
\be
C = \frac{\lambda^3 N}{4\pi^2}
\ee
and the {\it normalized sample size}
\be
\rho = \frac{\pi l}{\lambda}
\ee

\noindent as well as the quantities
\beq
R &=& \g C\rho \cdot Re[\frac{\hbar^2}{\wp^2 N}P^{(1)ret}]\\
\tilde{\rho} &=& 2\rho-\g C\rho \cdot Im[\frac{\hbar^2}{\wp^2 N}P^{(1)ret}]\\
Q&=&R+i\tilde{\rho}
\eeq
The result for \eqs{eq:G_Gb} then reads
\begin{widetext}
\begin{subequations}
\begin{align}
A_1& = \g^2 C\rho \frac{P_1}{R}\big(e^{R}-1\big)\\
A_2& =\frac{6\g^2 C \rho P_1}{Q^3}\bigg\{\frac{8Q}{(Q^*)^2}+\frac{e^{Q/2}(Q-2)(\tilde{\rho}e^{R}+iQ)}{R\tilde{\rho}}-\frac{ie^{Q^*/2}[R^2(R-2)+(R+10)\tilde{\rho}^2+3i\tilde{\rho}^3+iR\tilde{\rho}(3R-4)]}{\tilde{\rho}(Q^*)^2}\bigg\}\\
B& =32 \g^2 C^2 \rho^4 \frac{ P_2}{|Q|^4} \bigg|e^{Q^*/2}  \big(1-\frac{Q^*}{2}\big)\!-\!1\bigg|^2\approx B'
\end{align}
\end{subequations}
\end{widetext}
where $P_{1(2)}=\frac{\hbar^2 P^{(1,2)s}}{2 \wp^2 N^{1(2)}}$ are the source functions without prefactor. In order to obtain the symmetric form of $A_2$ with respect to the exchange of $r_1$ and $r_2$, one needs to add its complex conjugate and take the average. This amounts to taking the real part of the $A_2$ expression. Note that  $P_1$, $P_2$ and $P^{(1)ret}$ depend on the cooperative decay rate $\Gamma$. Therefore, $\G$ is obtained by solving the implicit function
\be
\label{eq:G_implicit}
\G=A_1(\G,a,\rho_{eg}...) + B(\G,a,\rho_{eg}...)
\ee
Notably, the collective decay rate $\Gamma$ primarily depend on the optical depth
\be
	\eta = C\rho = \frac{N\lambda^2 l}{4\pi}
\ee
which will ultimately determine the behavior of the system's dynamics. On the other hand, the collective light shift $\Delta$ is obtained by solving \eq{eq:light_shift} where we should replace $\w$ by $2\Delta+\delta$ according to the definition of $\w= \omega_0-\omega_l$ as the total light shift.

\section{Example: Homogeneous Gas of Two-Level Atoms}

\subsection{Cooperative Decay of the System}

As an example, we consider a homogeneous gas of two-levels atoms. We neglect the retardation effects of the light field, such that the atomic variables throughout the system changes simultaneously. This is a good approximation as long as the propagation time of the field $\tau_{prop}\!=\!l/c$ is much shorter than the timescale of cooperative decay $\tau_{coop}\!\sim\! \G^{-1}$. Therefore, the spatial dependence of the atomic variables in \eq{eq:nine_eqs} can be dropped. Note that this doesn’t mean the system is limited to zero size, since the calculation of cooperative decay rates in \eq{eq:G_Gb} involves an integration over an extended volume. For a non-driven system ($\Omega\!=\!0$), \eqs{eq:nine_eqs} are reduced to
\begin{subequations}
\begin{align}
\dot{a} &= -(2\G + \g)a + \G\\
\dot{n} &= -2(2\G + \g)n - 2\g(2a-1) + 8\bar{\G}x\\
\dot{x} &= -(2\G+\g)x + \bar{\G}n
\end{align}
\end{subequations}
with all other variables equaled to zero. We consider the system initially fully inverted, such that $a(0)=1$ and $n(0)=1$, and no coherences are present, i.e. $x(0)=0$. For simplicity, we neglect the induced light shift $\Delta$ and the spontaneous light shift $\delta$, and will solve \eq{eq:G_implicit} under the condition $\Delta\!=\!\delta\!=\!0$. This approximation is verified in Ref.~\cite{gray2016} for a non-driven system.
\begin{figure}
	\includegraphics[width=\linewidth]{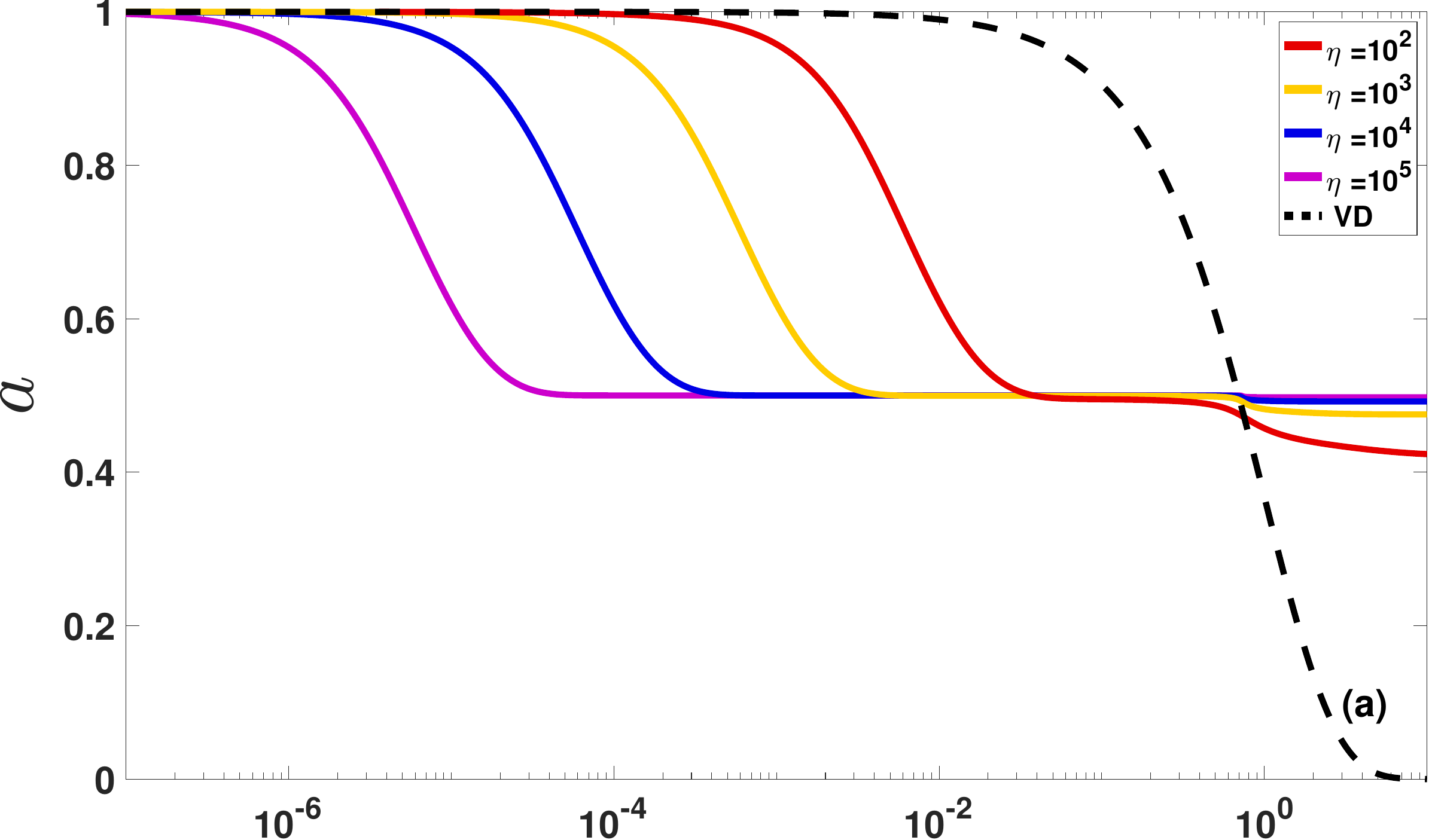}
	\includegraphics[width=\linewidth]{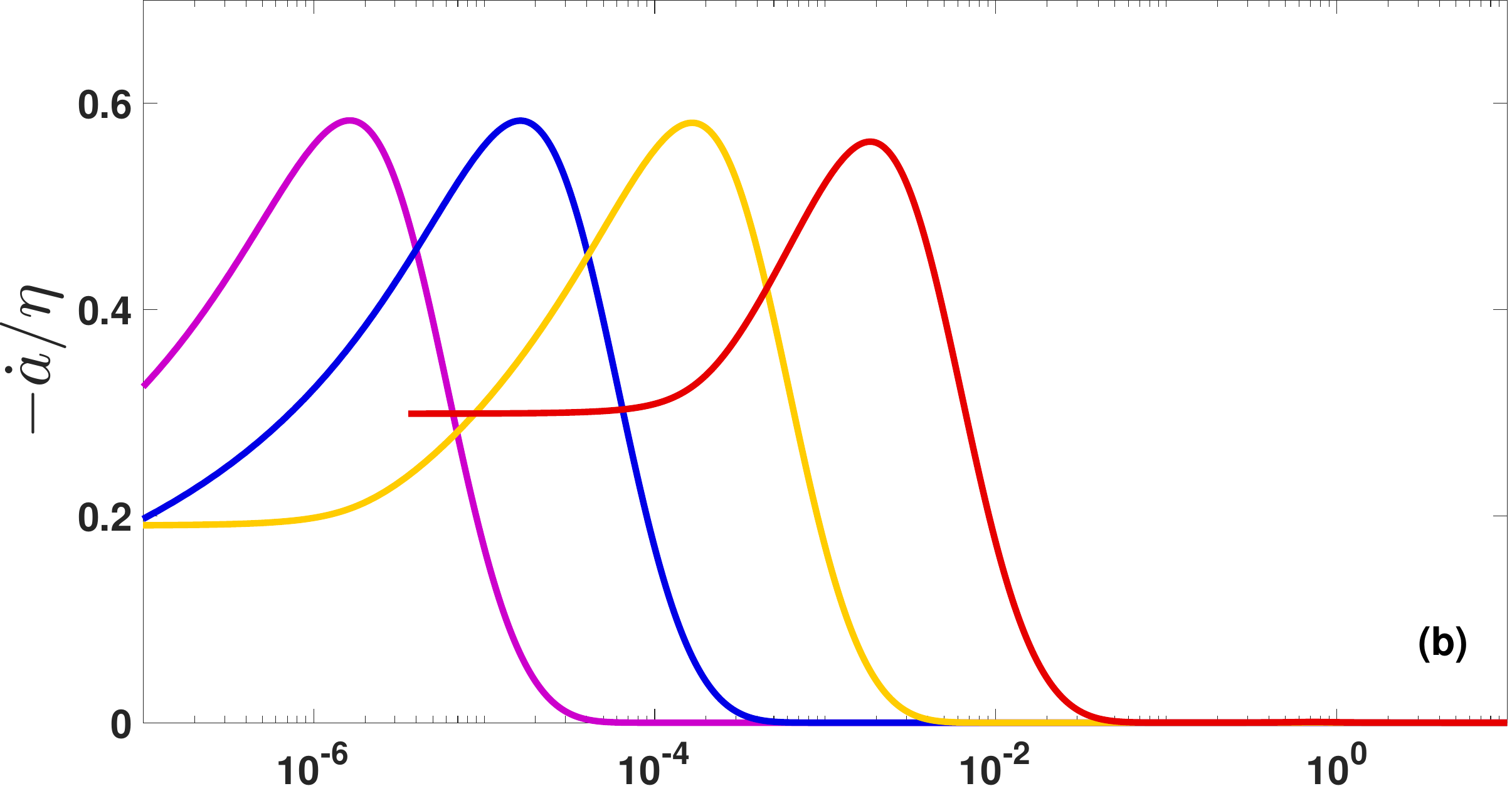}
	\includegraphics[width=\linewidth]{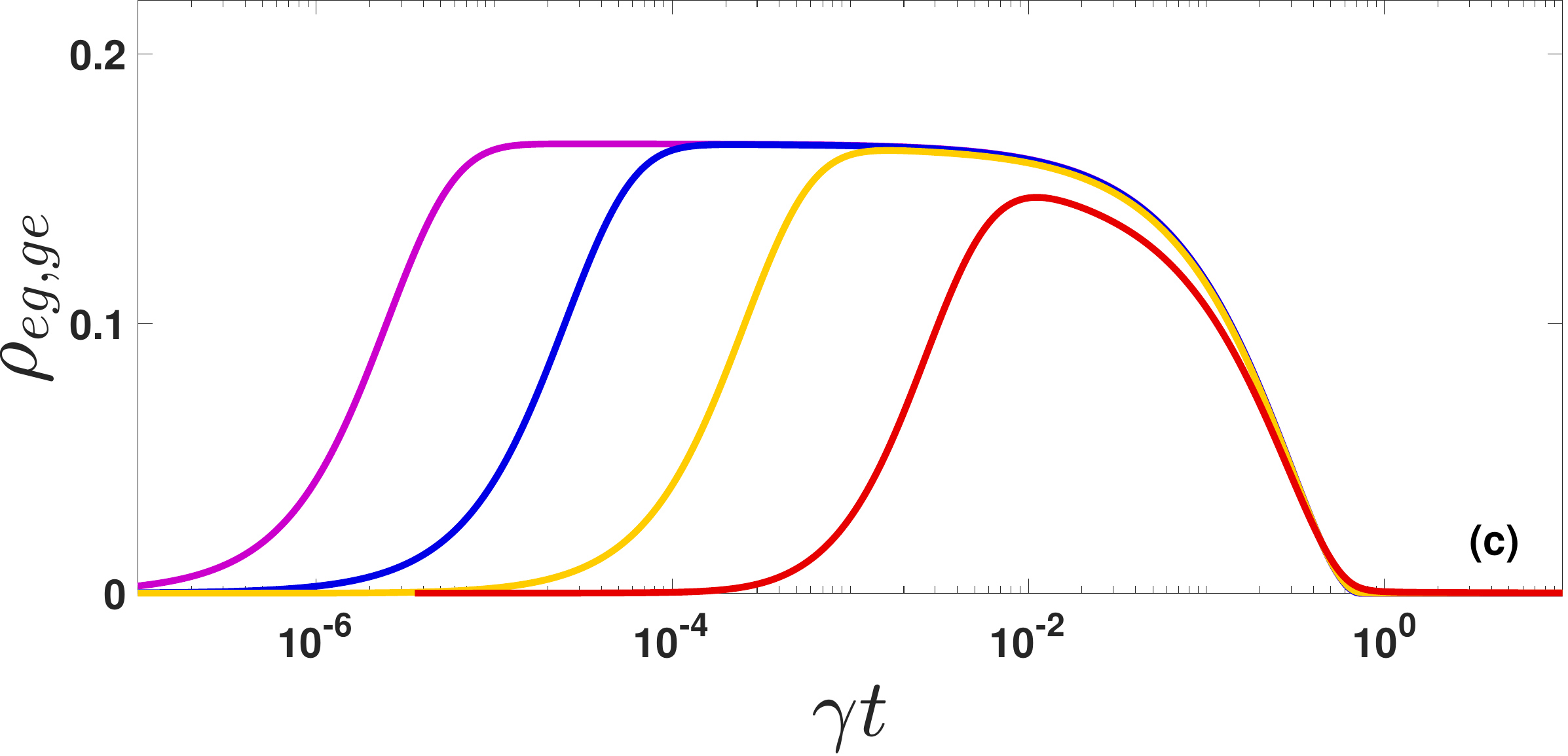}
	\caption{Numerical results showing super- and subradiance. (a) Excited state population for different optical depths $\eta$. Vacuum spontaneous decay dynamics is shown for comparison (dashed black line). (b) Rate of loss of population (proportional to the radiated intensity per atom) normalized by optical depth $\eta$. (c) Two-atom coherence $\rho_{egge}=\langle\sigma_1^+\sigma_2^-\rangle$. Time is given in the unit of vacuum life time $\gamma^{-1}$.}
	\label{fig:4}
\end{figure}

A simulation of the time-evolution is shown in \fig{fig:4}, plotted with different optical depth $\eta$. Note that the chosen optical depths cover several orders of magnitude, ranging from the typical value for a cold-atom experiment ($\sim\!\!10^2$)\cite{guerin2016}, to the value that can be reached by Rydberg-state superradiance ($\sim\!\!10^5$)\cite{Experiment_dense_7}. \fig{fig:4}(a) shows the time-evolution of the {\it average excited state population} $a$. As a comparison, the vacuum spontaneous decay for independent atoms (VD) is plotted with the black dashed line. Note that the time axis is in logarithmic scale and in units of $1/\gamma$. We observe an early decay where the excited population rapidly drops from $1$ to $\sim\!1/2$ compared to the vacuum spontaneous decay. This rapid decay corresponds to the superradiant outburst, as shown by the peaks in the radiated intensity per atom plotted in \fig{fig:4}(b). From \fig{fig:4}(b), the peaks of radiated intensity per atom ($-\dot{a}$) is proportional to the optical depth $\eta$, so the total intensity scales as ${\cal O}(N^2)$. Also, the times at which the maximum intensity occurs have an $N^{-1}$ dependence. After this initial superradiant phase, the radiation is suppressed at $a\approx 1/2$ and the system enters a subradiant regime. \fig{fig:4}(c) shows the coherence term $\rho_{eg,ge}$ from the two-atom density matrix. Its non-zero value that covers both the short-lived outburst and the long-lived suppressed emission verifies the cooperative nature of both phenomena. Starting from $\g t\approx 1$ where the coherence vanishes, the decay is enhanced again but attains only a relatively small value. This last phase of the time-evolution corresponds to {\it radiation trapping} \cite{holstein1947,RadiationTrapping_2}.

\begin{figure}
	\includegraphics[width=\linewidth]{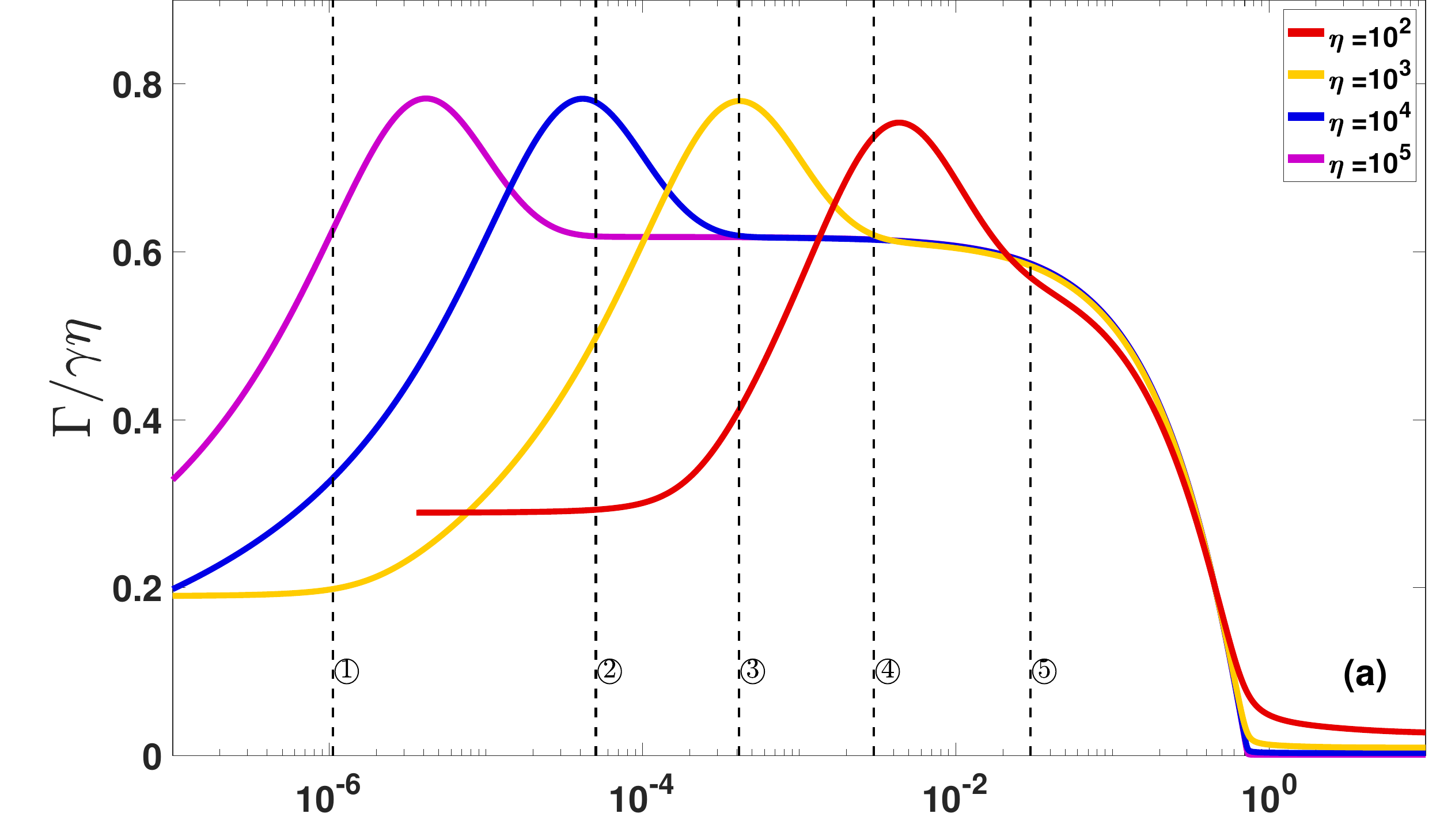}
	\includegraphics[width=\linewidth]{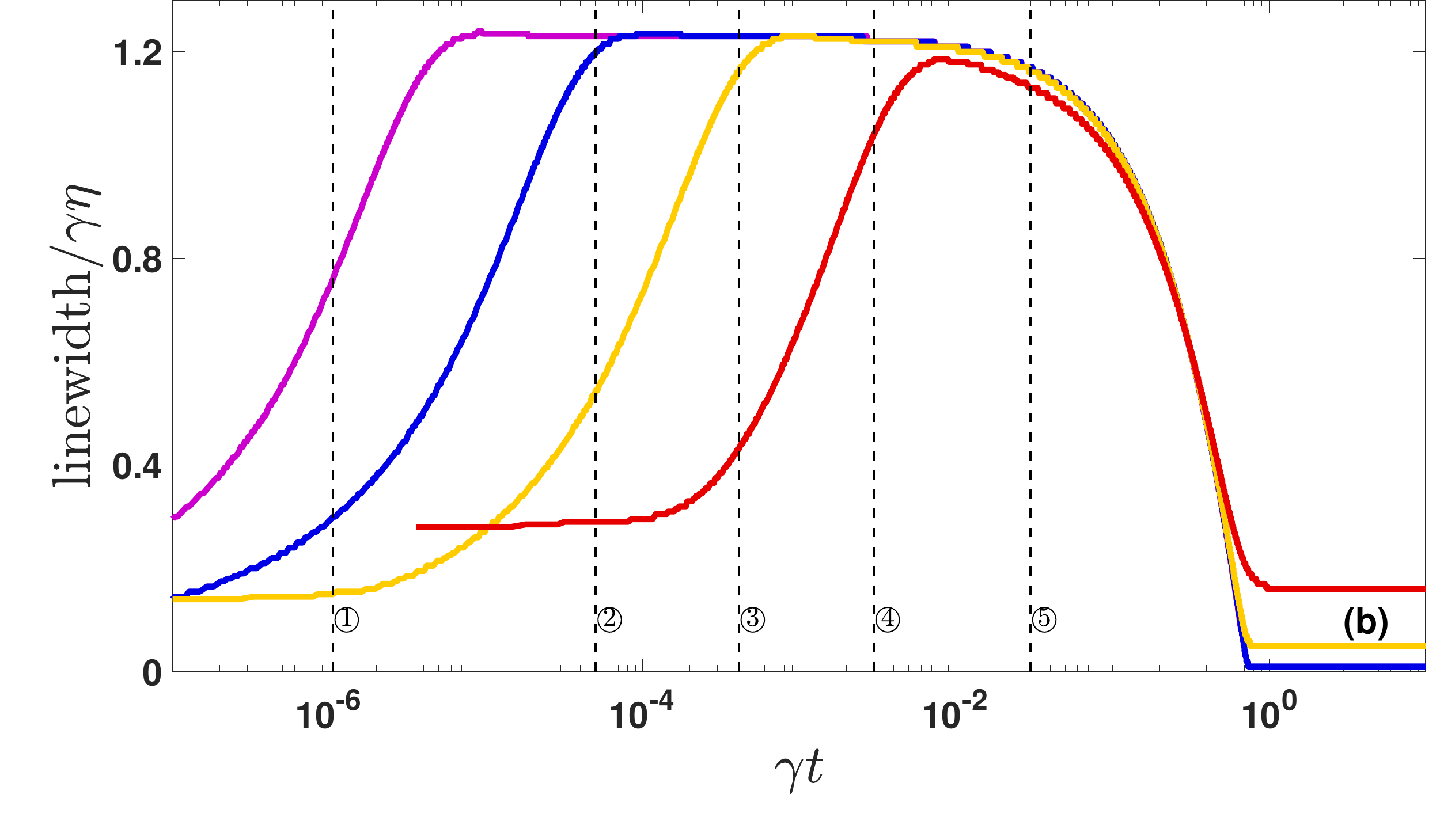}
	\includegraphics[width=\linewidth]{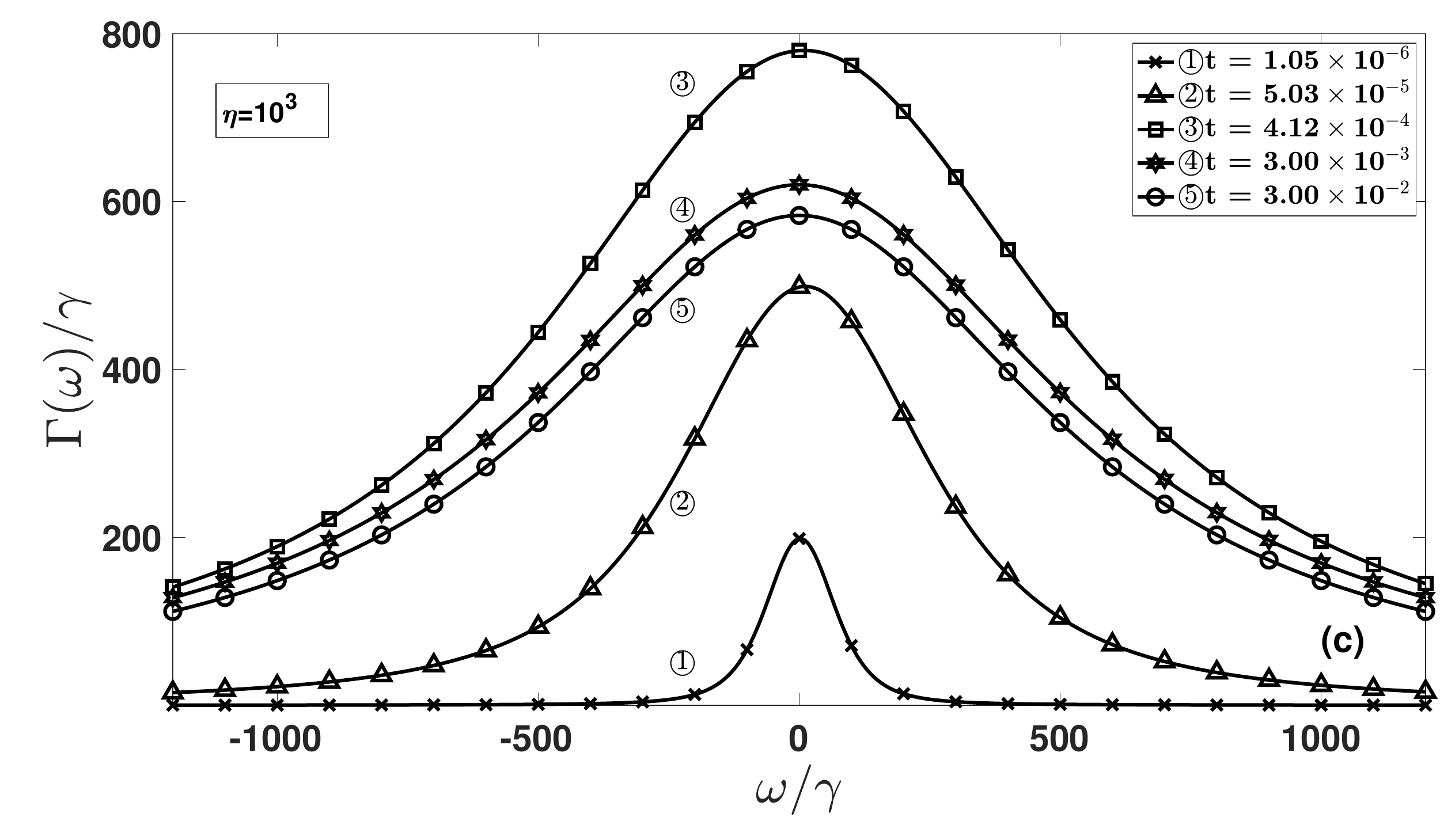}
	\caption{(a) Instantaneous cooperative decay rate $\G$ normalized by optical depth $\eta$. (b) Line width of the radiated spectrum, normalized by optical depth. (c) The spectrum of $\G$ at different times for $\eta=1000$. Numbered lines are plotted for times indicated in (a) and (b). Note that for all the plots $\G$ is in units of $\g$.}
	\label{fig:5}
\end{figure}

\fig{fig:5}(a) shows the time dependence of the approximated {\it cooperative decay rate} $\G(\w\!=\!0,t)$, by which the cooperative dynamic is mostly determined. Values are plotted in the unit of vacuum decay rate $\gamma$. $\G$ builds up to its maximum during the superradiance phase, remains nearly constant during the subradiance phase and finally vanishes at the radiation trapping regime. Since in \fig{fig:5}(a) the $\G$’s are normalized by the optical depth $\eta$, we see that the actual $\G$’s are indeed two to five orders of magnitude greater than the vacuum decay rate $\g$. Therefore, $\G$ has the major contribution to the overall radiated intensity, and possesses an ${\cal O}(N)$ dependence.

In \eqs{eq:cumulants} and \eq{eq:GG1}, $\G(\w,t)$ is expressed as the Fourier transform of the two-time field correlation, which, by virtue of the {\it optical Wiener-Khinchin theorem} \cite{meystre}, is related to the power spectral density of the field. The theorem is applicable to stationary processes, which is a well justified approximation as long as the correlating time scale $|\Delta\tau|$ is significantly shorter than the system evolution time scale $|\Delta t|$. Therefore, $\G(\w,t)$ is identical to the {\it instantaneous} spectrum of the cooperative radiation at time $t$. \fig{fig:5}(c) shows the instantaneous spectra at different times for a sample with optical depth $\eta=10^3$, whereas \fig{fig:5}(b) depicts the {\it instantaneous} linewidths, defined as the full width at half maximum. The circled numbers in \fig{fig:5}(c) corresponds to the vertical dashed lines in \fig{fig:5}(a,b), which represent different times. Note that the linewidths are broadened as the system evolves into the superradiant phase, and remain constant during the rest of the cooperative decay. The maximum linewidths scale with the optical depth $\eta$. It is worth noting that \fig{fig:4} and \fig{fig:5} exhibit a universal behavior of superradiant evolution with respect to changes in optical depth $\eta$.

\subsection{Subradiance}

As can be seen qualitatively in the time evolution of the decay (e.g. \fig{fig:4}), the dynamics of the system turns very slow once the quick superradiant outburst disappears. As we explain in this section, this phenomenon can be interpreted as a subradiant behavior. First, however, it is important to note that subradiance in fact emerges naturally in such a superradiant system. In Appendix~\ref{App:subradiant_states}, we demonstrate the transition from superradiant phase to subradiant phase in a generic multi-particle system, using the full Hilbert space consisting of both the symmetric manifold and non-symmetric manifolds. In this subsection, we  quantitatively investigate the subradiant phase of the homogeneous gas model. For that, we introduce the quantity $\xi\equiv -\dot{a}/a$, which represents {\it the number of emitted photons per excited atom per unit time}. For a constant $\xi (t)=\g_0$, the dynamic of the system reduces to an exponential decay $a=a_0\, \rm{exp}(-\g_0 t)$, where $\g_0$ is the inverse of the atomic lifetime $\tau_0$. A monotonically increasing (decreasing) $\xi(t)$ indicates a decay that is asymptotically faster (slower) than an exponential decay, which can occur in the transition between exponential decays with different instantaneous decay rates. To see this, we expand $\xi(t)$ to the first order around $t$, i.e. $\xi(t)=\xi_0 + \xi_1 t$, and integrate to obtain

\beq
a&=&a_0 e^{-\int \xi(t)dt} =a_0 e^{-\xi_1 t^2/2}e^{-\xi_0 t}\nonumber\\
&\equiv&a_0 \,g(t)\,e^{-\xi_0 t}
\eeq
For small $\xi_1$, the dynamics can be seen as a modified exponential decay with an amplitude $g(t)$. With $\xi_1<0$, $g(t)$ is monotonically increasing, which effectively describes a slowdown of the decay.

\begin{figure}
	\centering
	\includegraphics[width=\linewidth]{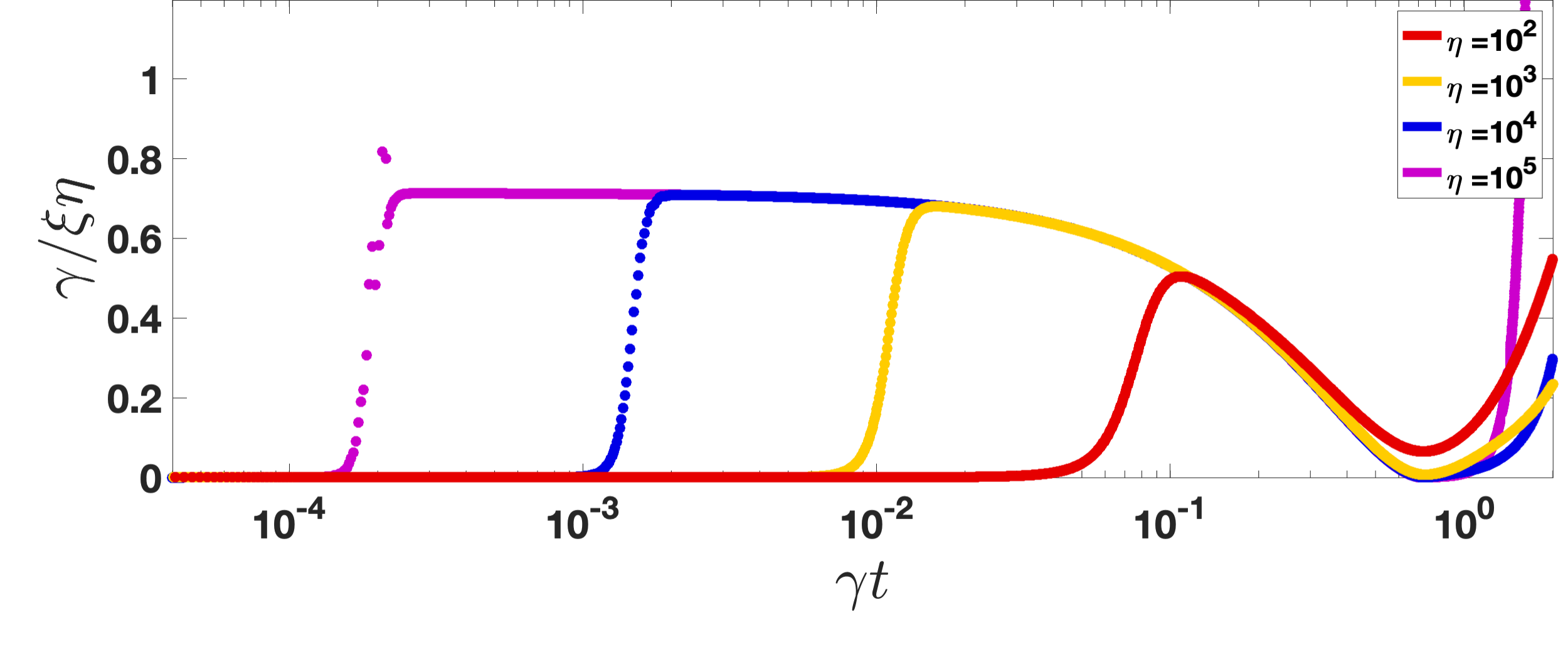}
	\caption{Study of the time-dependent nature of subradiance through the quantity $-a/(\eta\dot a) \equiv 1/\xi \eta$, plotted for different optical depths $\eta$. The constant plateau indicates an exponential decay, while the initial rise can be shown to be of polynomial time dependence.}
	\label{fig:a_over_adot}
\end{figure}

\fig{fig:a_over_adot} shows the inverse of $\xi \eta$ as a function of time. One can clearly see that $\xi$ remains constant during most of the subradiant phase. This is a signature of an exponential behavior featuring an extremely small instantaneous decay rate $\xi_0 \sim \gamma/0.7\eta$, orders of magnitude smaller than the vacuum decay rate $\g$. Optically dense media therefore exhibit a strong subradiant effect, with a characteristic lifetime $\tau_{{\rm sub}}\!\sim\! 1/\xi_0 \!\sim\! \eta/\gamma$ that scales linearly with the optical depth. These two features --- namely, the exponential behavior of subradiance and the scaling of its lifetime with the optical depth --- agree with recent experimental observations in clouds of cold atoms \cite{guerin2016}. Note also that the constant profile of $\xi$ finally disappears once the cooperative effect vanishes and radiation trapping starts to dominate.

\subsection{Phase Change in Radiation Field}

An experimental observation of a non-zero accumulated phase of the radiated field over time during superradiance in Rydberg-Rydberg transitions was reported in Ref.~\cite{Experiment_dense_7}. Here, we demonstrate that our method predicts this phase change during the superradiant phase.

From Eq.(\ref{eq:GG1}), $\G(\w,t)$ is expressed as the Fourier transform of the correlation function, up to a constant factor
\beq
\G(\w,t) \propto {\rm FT}\{\langle E^{-}(t)E^{+}(t+\Delta t)\rangle\}
\eeq
That is, the two-time correlation of the field operators can be simply obtained by its inverse Fourier transform
\be
\langle E^{-}(t)E^{+}(t+\Delta t)\rangle \propto \int_{-\infty}^{\infty} e^{-i\w \Delta t} \G(t,\w)d\w
\ee
The phase difference between the radiated field at different time is then just given by its complex phase angle
\be\label{eq:phase_angle}
\phi(t)\equiv-\rm{arg}(\langle E^{-}(t)E^{+}(t+\Delta t)\rangle)
\ee

This argument follows from the correspondence between classical and quantum correlation function, where the field operators are related to the positive and negative components of the classical field as $\hat{E}^\pm \!\sim\! {\cal E}^\pm$. In the rotating frame, one can assume a generic temporal relation of the classical field
\beq
{\cal E}^+(t+\Delta t) = e^{-i\phi(t)}{\cal E}^+(t)
\eeq
The quantum correlation function can then be interpreted as
\beq
\langle \hat{E}^{-}(t)\hat{E}^{+}(t+\Delta t)\rangle \sim \overline{|{\cal E}|^2 e^{-i\phi(t)}}
\eeq
where the over-line on the right hand side denotes the ensemble average. \eq{eq:phase_angle} is therefore verified.

\begin{figure}
\includegraphics[width=\linewidth]{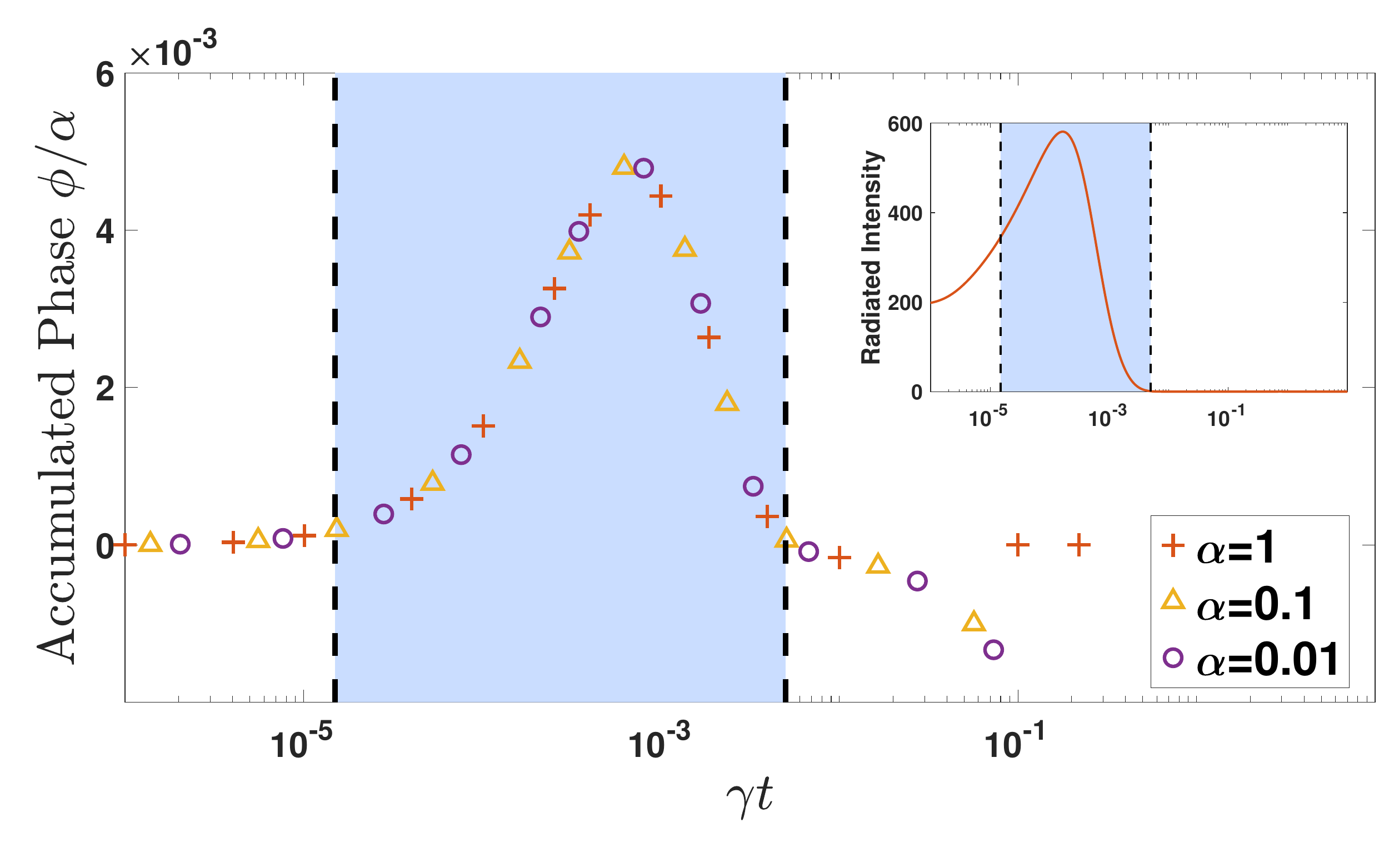}
\caption{Normalized accumulated phase obtained from the field correlation function, $\phi \equiv -\rm{arg}( \langle E^{-}(t)E^{+}(t+\Delta t)\rangle )$. The time differences are scaled according to the evolution time, $\Delta t=\alpha t$. Inset: radiated intensity showing a superradiant outburst. Shaded regions show the overlap between superradiance and the phase change.}
\label{fig:11}
\end{figure}

\fig{fig:11} presents the time evolution of the accumulated phase for an ensemble with optical depth $\eta=1000$, initially prepared in a fully inverted state. The inset displays the corresponding radiated intensity, where the peak signifies a superradiant outburst. The time differences are scaled according to the evolution time, $\Delta t=\alpha t$, where $\alpha$ is a scaling factor chosen from $1$, $0.1$ and $0.01$. The accumulated phase, normalized by $\alpha$, clearly reveals a robust phase evolution. The shaded regions in \fig{fig:11} highlight the overlap between superradiance and a non-zero phase change, which qualitatively aligns with the findings of Ref.~\cite{Experiment_dense_7}.

\section{Conclusions}

We have introduced an integrated method to describe an extended and optically dense ensemble of $N$ two-level atoms that radiate collectively. The dynamics of the system is formally derived with the help of the Keldysh formalism. By tracing out the quantized field and $(N-2)$ atoms, a two-atom effective description is obtained. The two-atom representation vastly reduces the difficulty of the many-body problem, while still capturing the essential collective nature of the full system. We established a self-consistent formalism that is satisfied at each physical time, and is thus able to solve the full time-evolution of the two-atom system. For a homogeneous gas of inverted two-level atoms, we demonstrate that the cooperative phenomena -- both superradiance and subradiance -- successively occur during the decay. These phenomena are characterized by a non-zero coherence $\rho_{eg,ge}$. The early-time superradiance features an outburst of radiated intensity that is proportional to the optical depth of the sample, while the cooperative decay rate is found to be orders of magnitude greater than the spontaneous decay rate of independent atoms. The subradiant phase, on the other hand, exhibits a slow exponential decay, having a decay rate that is inversely proportional to the optical depth, and a lifetime that scales linearly with the optical depth. Additionally, we find a broadened linewidth and a phase change of the radiated field during the cooperative decay, which qualitatively match with recent experimental observations of superradiant decay \cite{Experiment_dense_7}. At late times, when the coherence vanishes, the system enters a radiation trapping regime. These results therefore present our formalism as a powerful method to study cooperative effects in radiative many-body systems.

The formalism presented in this letter can also be leveraged to study optically dense systems in the presence of an external classical driving field and may elucidate the nature of the resulting steady states, which are of great interest due to their potential to exhibit bistability and spin-squeezing effects \cite{yelin1997,elie2014}. Additionally, this method can be readily employed to investigate the cooperative decay of systems with other geometries such as three-dimensional and two-dimensional ordered arrays of atoms \cite{rubiesbigorda2021superradiance}. Alternatively, lifting some symmetry constraints considered in this work could allow to study finite size effects of large samples.

\begin{acknowledgments}
This work has been supported by the NSF via PHY-2207972, and the CUA PFC PHY-2317134. ORB acknowledges support from Fundació Bancaria “la Caixa” (LCF/BQ/AA18/11680093). 
\end{acknowledgments}

\appendix

\section{Comparison with coupled-dipole model: Orders of correlations\label{app:A}}

\renewcommand{\thefigure}{A\arabic{figure}}
\setcounter{figure}{0}

\begin{figure}[h]
    \includegraphics[width=0.9\linewidth]{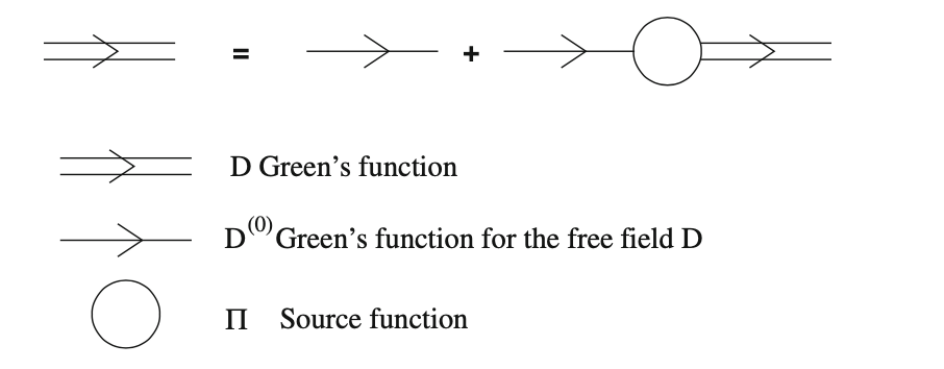}
    \caption{Graphical representation of the Dyson equation.}
    \label{fig:dyson}
\end{figure}

One of the most common theoretical methods for treating superradiant or cooperative systems is the \textit{coupled-dipole model}. In this model, the dynamics of an ensemble of $N$ two-level atoms is described by $2^N$ equations of motion that include a classical driving field as well as light-induced coherent and dissipative interactions between all pairs of emitters. These interactions are obtained from the real and imaginary parts of the Green's function, respectively. The mean field approximation is the simplest and most frequently used approximation, and relies on neglecting all correlations between emitters. When higher-order effects are considered, they are typically included using a (e.g. second-order) cumulant expansion \cite{kubo,cumulant_expansion_1,cumulant_expansion_2,cumulant_expansion_3}. Superficially, this method may seem similar to the one employed in this article. However, there is a crucial difference: we incorporate the effect of (two-body) correlations while accounting for multiple scattering to all orders. This is achieved through a self-consistent treatment in the calculation of the cumulants. The cumulants are calculated via the general Dyson equation (\eq{eq:dyson})
\begin{eqnarray}
D_{1\mu,2\nu}(\check{1},\check{2}) &=& D_{1\mu,2\nu}^{(0)}(\check{1},\check{2}) - \sum_{\alpha,\beta} \int \int d\check{1}' d\check{2}' D_{1\mu,1'\alpha}^{(0)}(\check{1},\check{1}') \nonumber\\
&&\times\Pi_{1'\alpha,2'\beta}(\check{1}',\check{2}') D_{2'\beta,2\nu}(\check{2}',\check{2}),
\end{eqnarray}
which has a graphical representation as shown in Fig.~\ref{fig:dyson}. The second term on the right-hand side includes the full expression via the term $D_{2'\beta,2\nu}(\check{2}',\check{2})$, leading to an infinite series that describes multiple scattering of all orders. This term does not appear in the second-order version of the coupled-dipole method. An analytic solution to the Dyson equation can often be obtained through a Fourier transform (see, e.g., Ref.~\cite{ms1999} and the appendix therein).

\section{Beyond coupled-dipole model: Distinction between spontaneous and induced effects\label{app:B}}

\renewcommand{\thefigure}{B\arabic{figure}}
\setcounter{figure}{0}

Cooperativity induces modifications in both the decay rates and frequencies of the transitions. These effects are commonly treated as overall collective phenomena, often referred to as {\it superradiant decay} or broadening, and the {\it collective Lamb shift}. However, it is crucial to distinguish between field intensity-dependent and spontaneous effects in many contexts. In non-cooperative settings, these would correspond to induced versus spontaneous emission, and the AC Stark shift versus the Lamb shift. Similarly, in this work, we differentiate between induced (intensity-dependent) and spontaneous (intensity-independent) effects. Notably, in the case of superradiance, the field or light intensity arises from the emission of photons from previously excited radiators.

The clear differentiation between these two types of effects becomes apparent once the analytic expressions for decay rates and frequency shifts are derived using our method (see the separate expressions in \eqs{eq:four_parameters}).

Specifically, the spontaneous decay rate and the free-space Lamb shift, represented by the $\gamma_{ij}$ and $\delta_{ij}$ terms in \eq{eq:master_eq} of the main text, are modified by the emission and re-absorption of virtual photons among many radiators. This process is independent of the total excitation present in the system. This independence is evident from the explicit expressions for $\gamma_{ij}$ and $\delta_{ij}$, which are proportional to the two-point correlator of the commutator $\sbrac{[E^+_j,E^-_i]} \sim \sbrac{[a,a^+]}$, and therefore do not depend on the number of photons.

Conversely, the superradiant decay rate $\Gamma_{ij}$ and the corresponding frequency shift $\Delta_{ij}$ are proportional to the cumulant $\dbrac{E^-_i E^+_j} \sim \sbrac{a^\dagger a}$, thus depending on the number of photons present in the ensemble, which evolves over time.

This distinction is further supported by Ref.~\cite{ma2024} and Section~II.C in this article, which demonstrates that the first type of effect depends only on the number density of atoms, while the second type depends on the optical depth.

\section{Keldysh Formalism\cite{keldysh}\label{app:keldysh}}

\renewcommand{\thefigure}{C\arabic{figure}}
\setcounter{figure}{0}

Suppose we have a Hamiltonian separated into free and interaction terms.
\[
	H \;=\; H_0 + V
\]
Recall the time-evolution operators in the Schrodinger and the interaction pictures, which are assigned with a time-ordering operator.
\begin{subequations}
\begin{align}
	U(t,t_0) &= \T e^{-\frac{i}{\hbar}\int_{t_0}^t d\tau H(\tau)}\\
	U_0(t,t_0) &= \T e^{-\frac{i}{\hbar}\int_{t_0}^t d\tau H_0(\tau)}\\
	U_I(t,t_0) &= U_0^\dagger(t,t_0) U(t,t_0) =  \T e^{-\frac{i}{\hbar}\int_{t_0}^t d\tau V^I(\tau)}
\end{align}
\end{subequations}
where $V^I(t) = U_0^\dagger(t,t_0) V(t) U_0(t,t_0)$ is the interaction term in the interaction picture. We consider $t>t_0$ in all cases. The key of Keldysh formalism is that the expectation value of any operator can alternatively be written in terms of the above time-evolution operators
\beq
	\langle\hat{O}\rangle(t) &=& {\rm Tr} \pare{\rho(t) \hat{O}}\nonumber\\
	&=& {\rm Tr} \pare{U(t,t_0) \rho_0 U^\dagger(t,t_0) \hat{O}}\nonumber\\
	&=& {\rm Tr} \pare{U(t,t_0) \rho_0 U^\dagger(t,t_0) U_0(t,t_0) \hat{O}^I(t) U_0^\dagger(t,t_0)}\nonumber\\
	&=& {\rm Tr} \pare{\rho_0 U^\dagger(t,t_0) U_0(t,t_0) \hat{O}^I(t) U_0^\dagger(t,t_0)U(t,t_0)}\nonumber\\
	&=& {\rm Tr} \pare{\rho_0 U_I^\dagger(t,t_0) \hat{O}^I(t) U_I(t,t_0)} \label{eq:end2}
\eeq

\begin{figure}[h]
	\includegraphics[width=\linewidth]{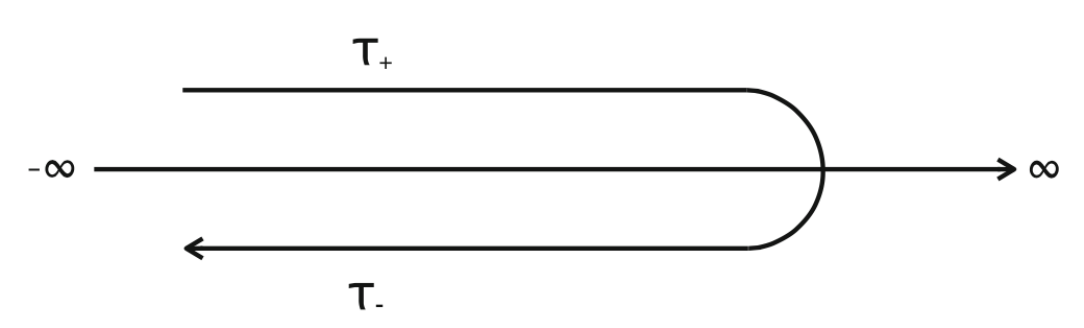}
	\caption{Schwinger-Keldysh contour in the time-evolution operator $S_{\cal C}$. The time integral starts from $-\infty$ on the upper branch, turns at present time $t$, and then goes back to $-\infty$ on the lower branch. Note that the time-ordering operator $\TC$ is defined differently on each branch: It's the ordinary time-ordering operator in the upper branch, and is the inverse time-ordering operator in the lower branch.}
	\label{fig:TC}
\end{figure}

The second step makes use of the time evolution of the density matrix in Schrodinger's picture. The third step expresses $\hat{O}$ in terms of the interaction picture operator $\hat{O}^I(t)$. While in the last step, we make use of the definition of $U_I(t,t_0)$. By defining the time-evolution operator along the Keldysh contour (see \fig{fig:TC})
\be
	S_{\cal C} \;\equiv\; {\cal T}_{\cal C} e^{-\frac{i}{\hbar}\int\limits_{\cal C} d\check\tau \,V^I(\check\tau)}
\ee
the expectation value of $\hat{O}$ can be written as
\be
	\langle\hat{O}\rangle(t) = {\rm Tr} \pare{\rho_0 \TC \SC \hat{O}^I(t)} \equiv \sbrac{\TC \SC \hat{O}^I(t)}_0
\ee
Note that the $\SC$ in this expression includes the two-fold $U_I^\dagger(t,t_0)$ and $U_I(t,t_0)$ in \eq{eq:end2}. Now that the matrix elements of the density operator are exactly the expectation value of the corresponding projection operators \cite{m1994}, i.e. $\rho_{\alpha\beta} = \sbrac{\ket{\beta}\!\!\bra{\alpha}}$, it is straightforward to write the density matrix elements in terms of $\SC$ and projection operators in the interaction picture.
\be
	\rho_{\alpha\beta}(t) = {\rm Tr} \pare{\rho_0 \TC \SC P_{\beta\alpha}^I(t)} \equiv \sbrac{\TC \SC P_{\beta\alpha}^I(t)}_0
\ee

\section{Cumulant Expansion for Effective Time-evolution Operator\label{app:cumulant}}

\renewcommand{\thefigure}{D\arabic{figure}}
\setcounter{figure}{0}

Suppose $X$, $Y$ and $Z$ are classical stochastic variables. The cumulant averages are defined as\cite{kubo}
\begin{eqnarray*}
	\dbrac{X}_1 &\equiv& \sbrac{X}\\
	\dbrac{XY}_2 &\equiv& \sbrac{XY}-\sbrac{X}\sbrac{Y}\\
	\dbrac{XYZ}_3 &\equiv& \sbrac{XYZ}-\sbrac{X}\sbrac{YZ}-\sbrac{Y}\sbrac{XZ}-\sbrac{Z}\sbrac{XY}\\
	&&+2\sbrac{X}\sbrac{Y}\sbrac{Z}\\
	&...&
\end{eqnarray*}
Comparing term by term, we readily obtain
\be
\sbrac{e^X} = \sbrac{{\rm exp}\left\{{\sum_{n=0}^\infty}\frac{1}{n!}X^n\right\}} = {\rm exp}\left\{{\sum_{n=1}^\infty}\frac{1}{n!}\dbrac{X^n}_n\right\}
\ee
This formula also works for operators, with the constrain that they mutually commute. This condition is satisfied with a time-ordering operator $\TC$ assigned to each term.

\section{\label{App:subradiant_states}Subradiant states}

\renewcommand{\thefigure}{E\arabic{figure}}
\setcounter{figure}{0}

\begin{figure}[h]
	\includegraphics[width=\linewidth]{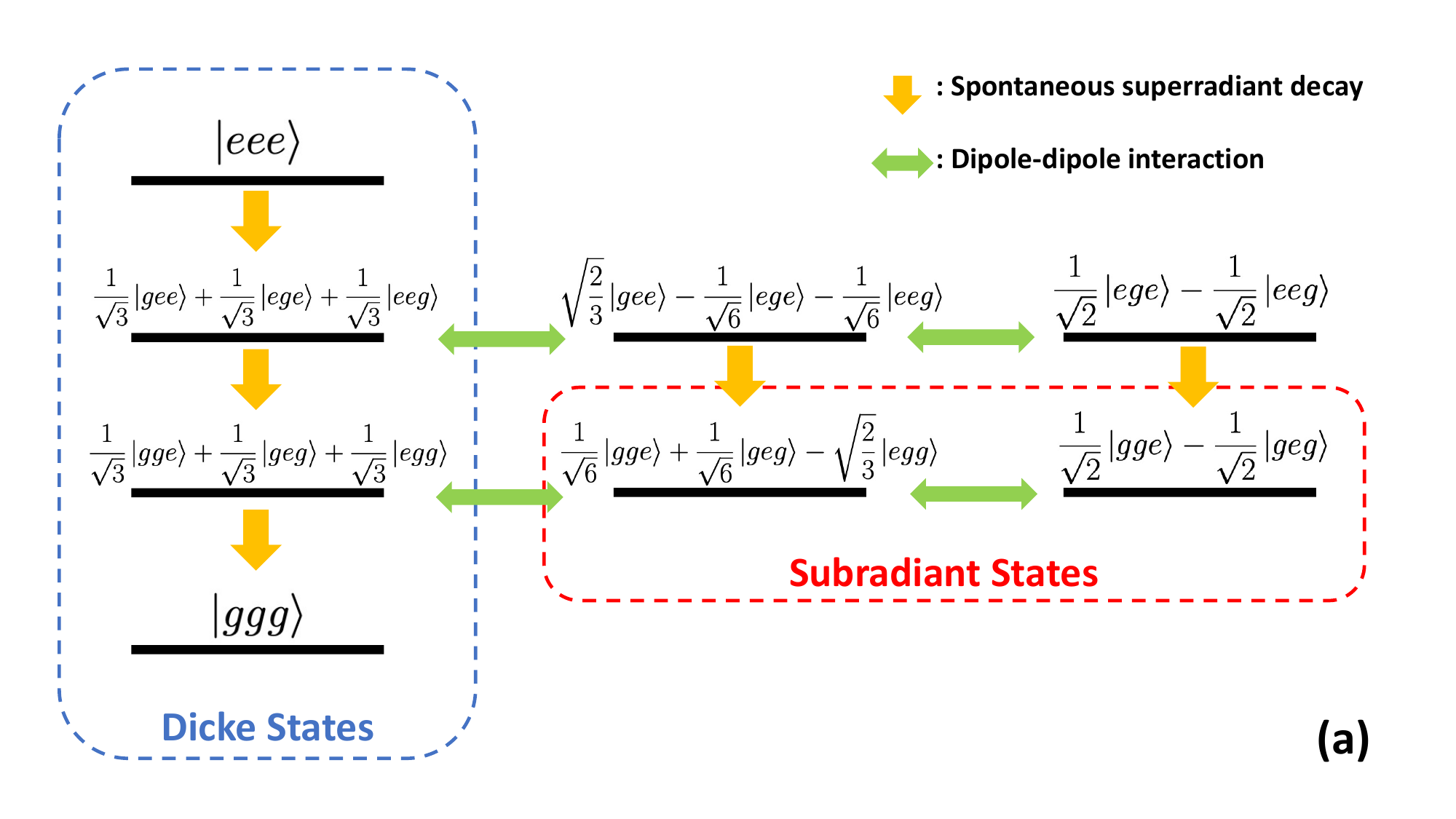}
	\includegraphics[width=\linewidth]{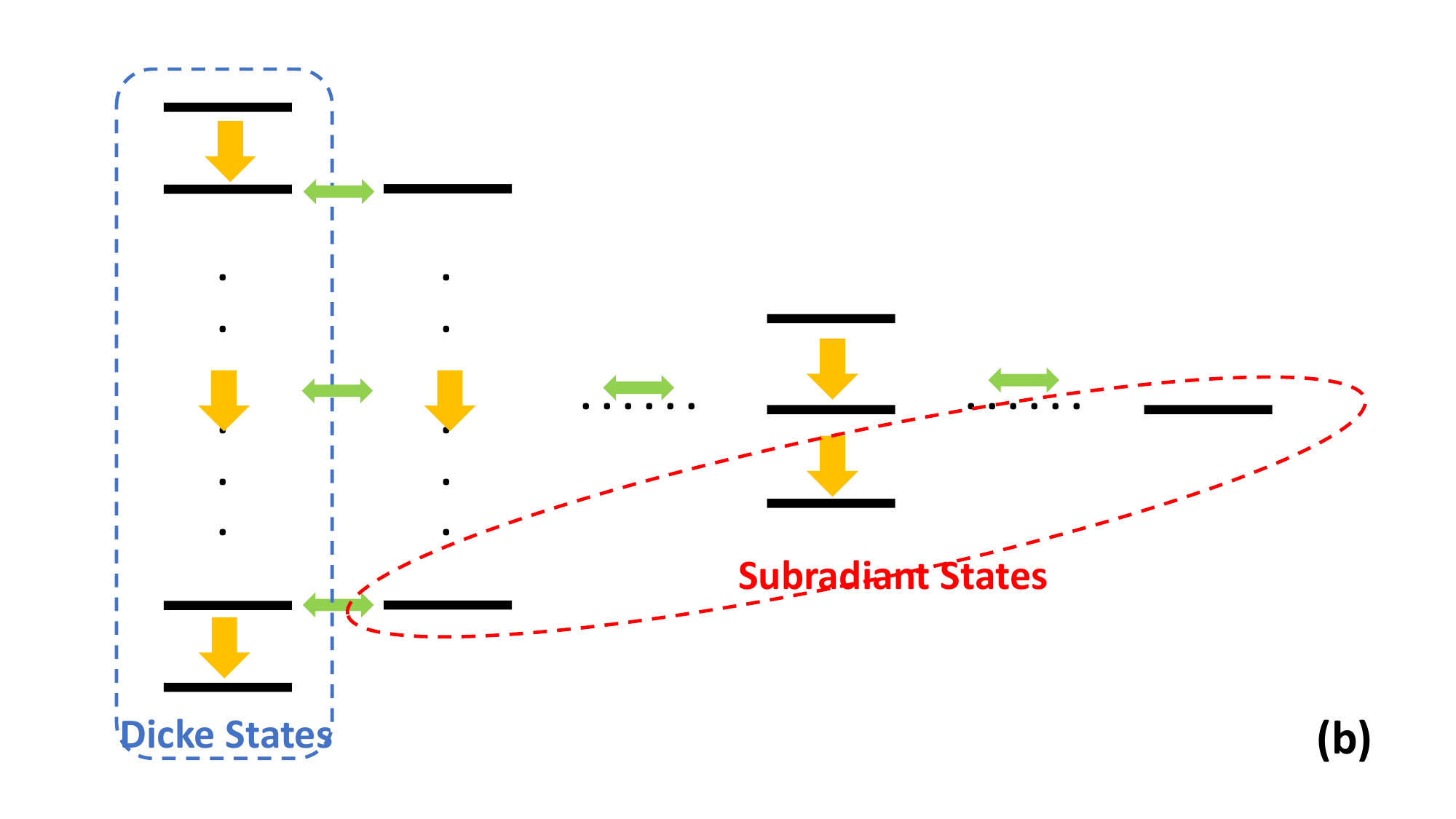}
	\caption{Schematic diagram of the cooperative radiation process for (a) three-particle system and (b) N-particle system.}
	\label{fig:subradiant_states}
\end{figure}

In \fig{fig:4}(a), the time-evolution of the initially inverted homogeneous gas exhibits a transition from the superradiant phase to the subradiant phase. This transition can be understood by means of the full manifolds of the Hilbert space, which includes both the symmetric Dicke states and the non-symmetric states. In \fig{fig:subradiant_states}, we demonstrate this process with the simple example of a three-particle system, as well as with a generic $N-$particle system.

Because of the broken permutational symmetry of finite-size systems, dipole-dipole interactions (represented by the green arrows) mediate transitions between the symmetric Dicke states and the non-symmetric states. Once the system evolves into one of the non-symmetric states, it cascades down the ``ladder'' through spontaneous superradiant decay (yellow arrows) until it reaches the subradiant states. There, the excitation remains trapped until the dipole-dipole interactions cause new transitions to other states that can radiate. This qualitatively explains the long lifetime of the subradiant plateau at $a\!\sim\! 1/2$ in \fig{fig:4}(a).

The non-symmetric states can be obtained from the symmetric Dicke states and the bare eigenstates of the free Hamiltonian by using the Gram-Schmidt orthogonalization method. Repeatedly applying the collective lowering operator $\sum_i \s_i$, one can keep track of the route of the decay path and finally reach the subradiant states.

\section{\label{AppC} Derivation of Master Equation}

\renewcommand{\thefigure}{F\arabic{figure}}
\setcounter{figure}{0}

We start with the effective time-evolution operator for the two atom system given in \eq{eq:Sceff}
\beqq
\dot \Se &=& \TC \frac{d}{dt}\bigg(\ih \int_{\cal C}d\tau_1 \sbrac{V^I(\tau_1)}\\
&&- \frac{1}{2\hbar^2} \iint_{\cal C}d\tau_1 d\tau_2 \dbrac{\TC V^I(\tau_1)V^I(\tau_2)}\bigg)e^P
\eeqq
The first term can be expanded as
\beqq
&&\frac{d}{dt}\pare{\ih \int_{\cal C}d\tau_1 \sbrac{V^I(\tau_1)}}  \\
&=& \ih\frac{d}{dt}\int_{\cal C}d\check\tau \sbrac{V^I(\check\tau)}\\
&=& \frac{i}{\hbar}\frac{d}{dt}\int_{\cal C}d\check\tau \sum_{i,\mu}\left(p^+_{i\mu}(\check\tau){\cal E}^-_{L,i\mu}(\check\tau) + p^-_{i\mu}(\check\tau){\cal E}^+_{L,i\mu}(\check\tau) \right)\\
&=& \frac{i}{\hbar} \sum_{i,\mu}\wp_\mu \left\{\left(\sigma_{i\mu}(t^+){\cal E}^-_{L,i\mu}(t^+) + \sigma_{i\mu}^\dagger(t^+){\cal E}^+_{L,i\mu}(t^+) \right) \right.\\
&&\left.- \left(\sigma_{i\mu}(t^-){\cal E}^-_{L,i\mu}(t^-) + \sigma_{i\mu}^\dagger(t^-){\cal E}^+_{L,i\mu}(t^-) \right) \right\}\\
\eeqq
Therefore, the first and second order terms of $\sbrac{\TC \dot\Se P^I_{\beta\alpha}(t)}$ are
\begin{widetext}
\beqq
&&FOT({\rm first\!-\!order\;\;terms}) = \sbrac{\TC \frac{d}{dt}\pare{\ih \int_{\cal C}d\tau_1 \sbrac{V^I(\tau_1)}} P^I_{\beta\alpha}(t) e^P}\\
&=& \frac{i}{\hbar} \sum_{i,\mu}\wp_\mu \sbrac{\TC  \left\{\sigma_{i\mu}(t^+){\cal E}^-_{L,i\mu}(t^+) + \sigma_{i\mu}^\dagger(t^+){\cal E}^+_{L,i\mu}(t^+) - \sigma_{i\mu}(t^-){\cal E}^-_{L,i\mu}(t^-) - \sigma_{i\mu}^\dagger(t^-){\cal E}^+_{L,i\mu}(t^-) \right\}P^I_{\beta\alpha}(t) \Se}\\
&=& \frac{i}{\hbar} \sum_{i,\mu}\wp_\mu {\rm Tr}\cur{\rho(t)\cur{P_{\beta\alpha}\sigma_{i\mu}{\cal E}^-_{L,i\mu} + P_{\beta\alpha}\sigma_{i\mu}^\dagger{\cal E}^+_{L,i\mu} - \sigma_{i\mu}{\cal E}^-_{L,i\mu}P_{\beta\alpha} - \sigma_{i\mu}^\dagger{\cal E}^+_{L,i\mu}P_{\beta\alpha}}}\\
&=& \frac{i}{\hbar} \sum_{i,\mu}\wp_\mu {\rm Tr}\cur{P_{\beta\alpha}\sigma_{i\mu}{\cal E}^-_{L,i\mu}\rho(t) + P_{\beta\alpha}\sigma_{i\mu}^\dagger{\cal E}^+_{L,i\mu}\rho(t) - \rho(t)\sigma_{i\mu}{\cal E}^-_{L,i\mu}P_{\beta\alpha} - \rho(t)\sigma_{i\mu}^\dagger{\cal E}^+_{L,i\mu}P_{\beta\alpha}}\\
&=& \frac{i}{\hbar} \sum_{i,\mu}\wp_\mu \bra{\alpha}\cur{\sigma_{i\mu}{\cal E}^-_{L,i\mu}\rho(t) + \sigma_{i\mu}^\dagger{\cal E}^+_{L,i\mu}\rho(t) - \rho(t)\sigma_{i\mu}{\cal E}^-_{L,i\mu}- \rho(t)\sigma_{i\mu}^\dagger{\cal E}^+_{L,i\mu}}\ket{\beta}\\
&=& \frac{i}{\hbar} \sum_{i,\mu}\wp_\mu \bra{\alpha}\squ{\sigma_{i\mu}{\cal E}^-_{L,i\mu} + \sigma_{i\mu}^\dagger{\cal E}^+_{L,i\mu},\rho(t)}\ket{\beta}\\
\eeqq
\beqq
&&SOT({\rm second\!-\!order\;\;terms}) = \sbrac{\TC ST \Se P^I_{\beta\alpha}(t)}\\
&=&\bigg\langle -\TC \sum_{i\mu,j\nu}\frac{\wp_\mu \wp_\nu}{2\hbar^2}\bigg(\\
&& \int_{-\infty}^t d\tau_2^+ \left\{\sigma_{i\mu}(t^+)D_{i\mu,j\nu}^{++}(t^+,\tau_2^+)\sigma_{j\nu}^\dagger(\tau_2^+)e^{-i\omega(t^+-\tau_2^+)} + \sigma_{i\mu}^\dagger(t^+)C_{i\mu,j\nu}^{++}(t^+,\tau_2^+)\sigma_{j\nu}(\tau_2^+)e^{i\omega(t^+-\tau_2^+)} \right\}\\
&+& \int_{-\infty}^t d\tau_1^+ \left\{\sigma_{i\mu}(\tau_1^+)D_{i\mu,j\nu}^{++}(\tau_1^+,t^+)\sigma_{j\nu}^\dagger(t^+)e^{-i\omega(\tau_1^+-t^+)} + \sigma_{i\mu}^\dagger(\tau_1^+)C_{i\mu,j\nu}^{++}(\tau_1^+,t^+)\sigma_{j\nu}(t^+)e^{i\omega(\tau_1^+-t^+)} \right\}\\
&+& \int_{-\infty}^t d\tau_2^- \left\{\sigma_{i\mu}(t^-)D_{i\mu,j\nu}^{--}(t^-,\tau_2^-)\sigma_{j\nu}^\dagger(\tau_2^-)e^{-i\omega(t^--\tau_2^-)} + \sigma_{i\mu}^\dagger(t^-)C_{i\mu,j\nu}^{--}(t^-,\tau_2^-)\sigma_{j\nu}(\tau_2^-)e^{i\omega(t^--\tau_2^-)} \right\}\\
&+& \int_{-\infty}^t d\tau_1^- \left\{\sigma_{i\mu}(\tau_1^-)D_{i\mu,j\nu}^{--}(\tau_1^-,t^-)\sigma_{j\nu}^\dagger(t^-)e^{-i\omega(\tau_1^--t^-)} + \sigma_{i\mu}^\dagger(\tau_1^-)C_{i\mu,j\nu}^{--}(\tau_1^-,t^-)\sigma_{j\nu}(t^-)e^{i\omega(\tau_1^--t^-)} \right\}\\
&-&\int_{-\infty}^t d\tau_2^- \left\{\sigma_{i\mu}(t^+)D_{i\mu,j\nu}^{+-}(t^+,\tau_2^-)\sigma_{j\nu}^\dagger(\tau_2^-)e^{-i\omega(t^+-\tau_2^-)} + \sigma_{i\mu}^\dagger(t^+)C_{i\mu,j\nu}^{+-}(t^+,\tau_2^-)\sigma_{j\nu}(\tau_2^-)e^{i\omega(t^+-\tau_2^-)} \right\}\\
&-& \int_{-\infty}^t d\tau_1^+ \left\{\sigma_{i\mu}(\tau_1^+)D_{i\mu,j\nu}^{+-}(\tau_1^+,t^-)\sigma_{j\nu}^\dagger(t^-)e^{-i\omega(\tau_1^+-t^-)} + \sigma_{i\mu}^\dagger(\tau_1^+)C_{i\mu,j\nu}^{+-}(\tau_1^+,t^-)\sigma_{j\nu}(t^-)e^{i\omega(\tau_1^+-t^-)} \right\}\\
&-& \int_{-\infty}^t d\tau_2^+ \left\{\sigma_{i\mu}(t^-)D_{i\mu,j\nu}^{-+}(t^-,\tau_2^+)\sigma_{j\nu}^\dagger(\tau_2^+)e^{-i\omega(t^--\tau_2^+)} + \sigma_{i\mu}^\dagger(t^-)C_{i\mu,j\nu}^{-+}(t^-,\tau_2^+)\sigma_{j\nu}(\tau_2^+)e^{i\omega(t^--\tau_2^+)} \right\}\\
&-& \int_{-\infty}^t d\tau_1^- \left\{\sigma_{i\mu}(\tau_1^-)D_{i\mu,j\nu}^{-+}(\tau_1^-,t^+)\sigma_{j\nu}^\dagger(t^+)e^{-i\omega(\tau_1^--t^+)} + \sigma_{i\mu}^\dagger(\tau_1^-)C_{i\mu,j\nu}^{-+}(\tau_1^-,t^+)\sigma_{j\nu}(t^+)e^{i\omega(\tau_1^--t^+)} \right\} \bigg)\\
&& \Se P^I_{\beta\alpha}(t) \bigg\rangle
\eeqq

There are in total 8 terms in $SOT$, here we show the calculation for the first term as an example.
\beqq
T_1 &=& \bigg\langle \TC \int_{-\infty}^t d\tau_2^+ \left\{\sigma_{i\mu}(t^+)D_{i\mu,j\nu}^{++}(t^+,\tau_2^+)\sigma_{j\nu}^\dagger(\tau_2^+)e^{-i\omega(t^+-\tau_2^+)} \right.\\
&& \left.+ \sigma_{i\mu}^\dagger(t^+)C_{i\mu,j\nu}^{++}(t^+,\tau_2^+)\sigma_{j\nu}(\tau_2^+)e^{i\omega(t^+-\tau_2^+)} \right\}  \Se P^I_{\beta\alpha}(t) \bigg\rangle\\
&=& \bigg\langle \TC \int_{-\infty}^t d\tau_2^+ \left\{P^I_{\beta\alpha}(t)\sigma_{i\mu}(t^+)\sigma_{j\nu}^\dagger(\tau_2^+)D_{i\mu,j\nu}^{++}(t^+,\tau_2^+)e^{-i\omega(t^+-\tau_2^+)} \right.\\
&& \left.+ P^I_{\beta\alpha}(t)\sigma_{i\mu}^\dagger(t^+)\sigma_{j\nu}(\tau_2^+)C_{i\mu,j\nu}^{++}(t^+,\tau_2^+)e^{i\omega(t^+-\tau_2^+)} \right\}  \Se  \bigg\rangle\\
&=& \bigg\langle \int_{-\infty}^t d\tau_2^+ \left\{P^H_{\beta\alpha}(t)\sigma_{i\mu}^H(t^+)\sigma_{j\nu}^{\dagger H}(\tau_2^+)D_{i\mu,j\nu}^{++}(t^+,\tau_2^+)e^{-i\omega(t^+-\tau_2^+)} \right.\\
&& \left.+ P^H_{\beta\alpha}(t)\sigma_{i\mu}^{\dagger H}(t^+)\sigma_{j\nu}^H(\tau_2^+)C_{i\mu,j\nu}^{++}(t^+,\tau_2^+)e^{i\omega(t^+-\tau_2^+)} \right\}  \bigg\rangle\\
&=& \bigg\langle \int_{-\infty}^t d\tau' \left\{P^H_{\beta\alpha}(t)\sigma_{i\mu}^H(t)\sigma_{j\nu}^{\dagger H}(\tau')D_{i\mu,j\nu}^{++}(t,\tau')e^{-i\omega(t-\tau')} \right.\\
&& \left.+ P^H_{\beta\alpha}(t)\sigma_{i\mu}^{\dagger H}(t)\sigma_{j\nu}^H(\tau')C_{i\mu,j\nu}^{++}(t,\tau')e^{i\omega(t-\tau')} \right\}  \bigg\rangle\\
&=& \bigg\langle \int_{-\infty}^t d\tau' \left\{P^H_{\beta\alpha}(t)\sigma_{i\mu}^H(t)\sigma_{j\nu}^{\dagger H}(\tau')\dbrac{E^-_{i\mu}(t)E^+_{j\nu}(\tau')}e^{-i\omega(t-\tau')} \right.\\
&& \left.+ P^H_{\beta\alpha}(t)\sigma_{i\mu}^{\dagger H}(t)\sigma_{j\nu}^H(\tau')\dbrac{E^+_{i\mu}(t)E^-_{j\nu}(\tau')}e^{i\omega(t-\tau')} \right\}  \bigg\rangle\\
&=& \bigg\langle \int^{+\infty}_0 d\tau \left\{P^H_{\beta\alpha}(t)\sigma_{i\mu}^H(t)\sigma_{j\nu}^{\dagger H}(t-\tau)\dbrac{E^-_{i\mu}(t)E^+_{j\nu}(t-\tau)}e^{-i\omega\tau} \right.\\
&& \left.+ P^H_{\beta\alpha}(t)\sigma_{i\mu}^{\dagger H}(t)\sigma_{j\nu}^H(t-\tau)\dbrac{E^+_{i\mu}(t)E^-_{j\nu}(t-\tau)}e^{i\omega\tau} \right\}  \bigg\rangle\\
&=& \bigg\langle \int^{+\infty}_0 d\tau \left\{P^H_{\beta\alpha}(t)\sigma_{i\mu}^H(t)\sigma_{j\nu}^{\dagger H}(t)\dbrac{E^-_{i\mu}(t)E^+_{j\nu}(t-\tau)}e^{-i\omega\tau} \right.\\
&& \left.+ P^H_{\beta\alpha}(t)\sigma_{i\mu}^{\dagger H}(t)\sigma_{j\nu}^H(t)\dbrac{E^+_{i\mu}(t+\tau)E^-_{j\nu}(t)}e^{i\omega\tau} \right\}  \bigg\rangle\\
&=& \bra{\alpha} \int^{+\infty}_0 d\tau \left\{\sigma_{i\mu}\sigma_{j\nu}^{\dagger}\rho(t)\dbrac{E^-_{i\mu}(t)E^+_{j\nu}(t-\tau)}e^{-i\omega\tau} + \sigma_{j\nu}^{\dagger}\sigma_{i\mu}\rho(t)\dbrac{E^+_{j\nu}(t+\tau)E^-_{i\mu}(t)}e^{i\omega\tau} \right\}  \ket{\beta}\\
\eeqq
\end{widetext}
In the 1st and 2nd steps all operators are in interaction picture, in the 4th step we remove the superscript on $t$ and replace $\tau_2$ by $\tau'$. In the 6th step we make a variable substitution $\tau = t-\tau'$. In the 7th step we make the Markov approximation, such that all atomic operators depend on t. For correlations with an associated timescale much smaller than that of the system evolution, we employ a quasi-stationary approximation such that $\dbrac{E^\mp_{i\mu}(t)E^\pm_{j\nu}(t-\tau)} \approx \dbrac{E^\mp_{i\mu}(t+\tau)E^\pm_{j\nu}(t)}$. In the last step we go back to Schrodinger's picture and reassign dummy indexes if needed.

Here we list the results of the remaining seven terms:
\begin{widetext}
\beqq
T_2 &=& \bra{\alpha} \int^{+\infty}_0 d\tau \left\{\sigma_{j\nu}^\dagger\sigma_{i\mu}\rho(t)\dbrac{E^+_{j\nu}(t+\tau)E^-_{i\mu}(t)}e^{i\omega\tau} + \sigma_{i\mu}\sigma_{j\nu}^\dagger\rho(t)\dbrac{E^-_{i\mu}(t)E^+_{j\nu}(t-\tau)}e^{-i\omega\tau} \right\} \ket{\beta}\\
T_3 &=& \bra{\alpha} \int^{+\infty}_0 d\tau \left\{\rho(t)\sigma_{j\nu}^\dagger\sigma_{i\mu}\dbrac{E^+_{j\nu}(t-\tau)E^-_{i\mu}(t)}e^{-i\omega\tau} + \rho(t)\sigma_{i\mu}\sigma_{j\nu}^\dagger\dbrac{E^-_{i\mu}(t)E^+_{j\nu}(t+\tau)}e^{i\omega\tau} \right\}  \ket{\beta}\\
T_4 &=& \bra{\alpha} \int^{+\infty}_0 d\tau \left\{\rho(t)\sigma_{i\mu}\sigma_{j\nu}^\dagger\dbrac{ E^-_{i\mu}(t)E^+_{j\nu}(t+\tau)}e^{i\omega\tau} + \rho(t)\sigma_{j\nu}^\dagger\sigma_{i\mu}\dbrac{E^+_{j\nu}(t-\tau)E^-_{i\mu}(t)}e^{-i\omega\tau} \right\} \ket{\beta}\\
T_5 &=& \bra{\alpha} \int^{+\infty}_0 d\tau \left\{\sigma_{i\mu}\rho(t)\sigma_{j\nu}^\dagger\dbrac{E^+_{j\nu}(t-\tau)E^-_{i\mu}(t)}e^{-i\omega\tau} + \sigma_{j\nu}^\dagger\rho(t)\sigma_{i\mu}\dbrac{E^-_{i\mu}(t)E^+_{j\nu}(t+\tau)}e^{i\omega\tau} \right\} \ket{\beta}\\
T_6 &=& \bra{\alpha} \int^{+\infty}_0 d\tau \left\{\sigma_{i\mu}\rho(t)\sigma_{j\nu}^\dagger\dbrac{E^+_{j\nu}(t+\tau)E^-_{i\mu}(t)}e^{i\omega\tau} + \sigma_{j\nu}^\dagger\rho(t)\sigma_{i\mu}\dbrac{E^-_{i\mu}(t)E^+_{j\nu}(t-\tau)}e^{-i\omega\tau} \right\} \ket{\beta}\\
T_7 &=& \bra{\alpha} \int^{+\infty}_0 d\tau \left\{\sigma_{j\nu}^\dagger\rho(t)\sigma_{i\mu}\dbrac{E^-_{i\mu}(t)E^+_{j\nu}(t-\tau)}e^{-i\omega\tau} + \sigma_{i\mu}\rho(t)\sigma_{j\nu}^\dagger\dbrac{E^+_{j\nu}(t+\tau)E^-_{i\mu}(t)}e^{i\omega\tau} \right\} \ket{\beta}\\
T_8 &=& \bra{\alpha} \int^{+\infty}_0 d\tau \left\{\sigma_{j\nu}^\dagger\rho(t)\sigma_{i\mu}\dbrac{E^-_{i\mu}(t)E^+_{j\nu}(t+\tau)}e^{i\omega\tau} + \sigma_{i\mu}\rho(t)\sigma_{j\nu}^\dagger\dbrac{E^+_{j\nu}(t-\tau)E^-_{i\mu}(t)}e^{-i\omega\tau} \right\} \ket{\beta}
\eeqq
\end{widetext}
with
\be
SOT = -\!\sum_{i\mu,j\nu}\frac{\wp_\mu \wp_\nu}{2\hbar^2} \pare{T_1 \!+\! T_2 \!+\! T_3 \!+\!T_4 \!-\! T_5 \!-\! T_6 \!-\! T_7 \!-\! T_8}
\ee

\section{\label{App:retarded_greens_function}Retarded Green's function in the medium}

\renewcommand{\thefigure}{G\arabic{figure}}
\setcounter{figure}{0}

According to Ref. \cite{ms1999}, the Fourier transform of the retarded Green's function in the atomic medium can be written as

\begin{equation}
\label{eq:dret_tensor}
    \mathbf{\tilde{\tilde{D}}^{ret}}(\vec{q},\omega_0,t)= \left[ \mathbf{1}+\mathbf{\tilde{\tilde{D}}_0^{ret}}(\vec{q},\omega_0) \cdot \mathbf{\tilde{P}}^{ret}(\omega_0,t) \right]^{-1} \cdot \mathbf{\tilde{\tilde{D}}_0^{ret}}(\vec{q},\omega_0)
\end{equation}

\noindent where $\mathbf{\tilde{P}}^{ret}(\omega_0,t)$ is the polarization function and $\mathbf{\tilde{\tilde{D}}_0^{ret}}(\vec{q},\omega_0)$ is the free-space retarded propagator. 

The three-dimensional free-space retarded propagator in real space is 
\begin{widetext}
\beq
\label{eq: free_space retarded propagator}
    D_{0\alpha\beta}^{ret}(\vec{x},\omega_0)&=& -\frac{i\hbar}{4\pi \epsilon_0} \left(\frac{\omega_0^2}{c^2}\delta_{\alpha\beta}+ \frac{\partial^2}{\partial x_{\alpha} \partial x_{\beta}} \right) \frac{e^{-i\omega_0 r/c}}{r} \nonumber \\
    &=& -\frac{i\hbar k_0^2}{4\pi \epsilon_0} \left[\left(1-\frac{i}{k_0 r} -\frac{1}{k_0^2r^2} \right) \delta_{\alpha\beta} +\left( -1+\frac{3i}{k_0r} + \frac{3}{k_0^2 r^2}\right) \frac{x_\alpha x_\beta}{r^2}\right] \frac{e^{-i k_0 r}}{r}
\eeq
\end{widetext}

\noindent with $r=|\vec{x}|$ and $k_0=\omega_0/c$.

For a randomly polarized medium we can apply the polarization average $\langle \mathcal{P}_\alpha \mathcal{P}_\beta \rangle = \frac{1}{3} \delta_{\alpha \beta}$, which is equivalent to performing the orientation average 

\begin{equation}
    \frac{x_\alpha x_\beta}{r^2} \rightarrow \langle \frac{x_\alpha x_\beta}{r^2} \rangle = \frac{1}{3} \delta_{\alpha \beta}
\end{equation}

Under these conditions, the free-space Green's function becomes a spherical tensor

\begin{eqnarray}
\mathbf{\tilde{D}}_0^{ret}(\vec{x},\omega_0) 
    = -\frac{i\hbar k_0^2}{6\pi \epsilon_0}  \frac{e^{-i k_0 r}}{r} \mathbf{1} 
\end{eqnarray}

\noindent and the corresponding Fourier transform can be written as

\beqq
    \mathbf{\tilde{\tilde{D}}_0^{ret}}(\vec{q},\omega_0)&=& \int_{V_\infty} d^3 \vec{x} \mathbf{\tilde{D}}_0^{ret}(\vec{x},\omega_0) e^{i\vec{q}\vec{x}} e^{-\epsilon r}\\
    &=&-\frac{2i\hbar k_0^2}{3\epsilon_0} \frac{1}{q^2-k_0^2+2i k_0 \epsilon} \mathbf{1}
\eeqq

\noindent with $q=|\vec{q}|$ and where we have introduced the small, positive constant $\epsilon$ that allows convergence of the infinite volume integral. 

The spherical nature of the Fourier transformed free-space propagator tensor turns Eq.(\ref{eq:dret_tensor}) into a scalar equation. The retarded propagator in the atomic medium in momentum space takes now the simple form 

\begin{equation}
    \tilde{\tilde{D}}^{ret} (\vec{q},\omega_0) = -\frac{2i\hbar k_0^2}{3\epsilon_0} \frac{1}{q^2-k_0^2 \left( 1+ \frac{2i\hbar}{3\epsilon_0} \tilde{P}^{ret}\right)+2i k_0 \epsilon}
\end{equation}

Going back to real space, we obtain

\beq
&&\tilde{D}^{ret} (\vec{x},\omega_0) = \frac{1}{(2\pi)^3} \int d^3 \vec{q} \tilde{\tilde{D}}^{ret} (\vec{q},\omega_0) e^{i \vec{q} \vec{x}} \\
&=& \frac{\hbar k_0^2}{6 \pi^2 \epsilon_0} \frac{1}{r} \int_{-\infty}^\infty dq \frac{q e^{-iqr}}{q^2-k_0^2 \left( 1+ \frac{2i\hbar}{3\epsilon_0} \tilde{P}^{ret} - \frac{2i \epsilon}{k_0}\right)}
\eeq

This integral is solved by contour integration along the real axis and a semicircle on the lower half plane, where $e^{-iqr} \rightarrow 0$ for large $|q|$. The integrand has two poles at $Re[q] \approx \pm k_0$. For an absorbing medium with $Re[\tilde{P}^{ret}]<0$, the pole at $Re[q] \approx k_0$ is in the lower half-plane, whereas the pole at $Re[q] \approx -k_0$ is in the lower one. The integral therefore results in a retarded propagator. Similarly, a retarded propagator is obtained for $Re[\tilde{P}^{ret}]=0$ by introducing a small, positive $\epsilon$. For an amplifying medium with $Re[\tilde{P}^{ret}]>0$, the pole at $Re[q] \approx k_0$ is in the upper half-plane. This problem arises from the fact that the retarded propagator of an amplifying medium increases  exponentially with distance and can therefore not be Fourier transformed. Instead, a cutoff function should be included to take into account the finite size of the medium. Here, we will assume that the effect of the cutoff is analogous to adding a sufficiently large $\epsilon$ that moves the pole to the lower half-plane (see Ref. \cite{ms1999}). 

Using the residue theorem, we finally find

\begin{equation}
    \tilde{D}^{ret} (\vec{x},\omega_0) = 
    -\frac{i\hbar k_0^2}{6 \pi \epsilon_0} \frac{e^{-i q_0 r}}{r}
\end{equation}

\noindent where 

\begin{equation}
\label{eq: q0}
    q_0= k_0 \sqrt{1+\frac{2i\hbar}{3\epsilon_0} \tilde{P}^{ret}}
\end{equation}

\section{\label{App:taylor_expansion}Taylor Expansion in the Retarded Green's Function}

\renewcommand{\thefigure}{H\arabic{figure}}
\setcounter{figure}{0}

\begin{figure}
\includegraphics[width=\linewidth]{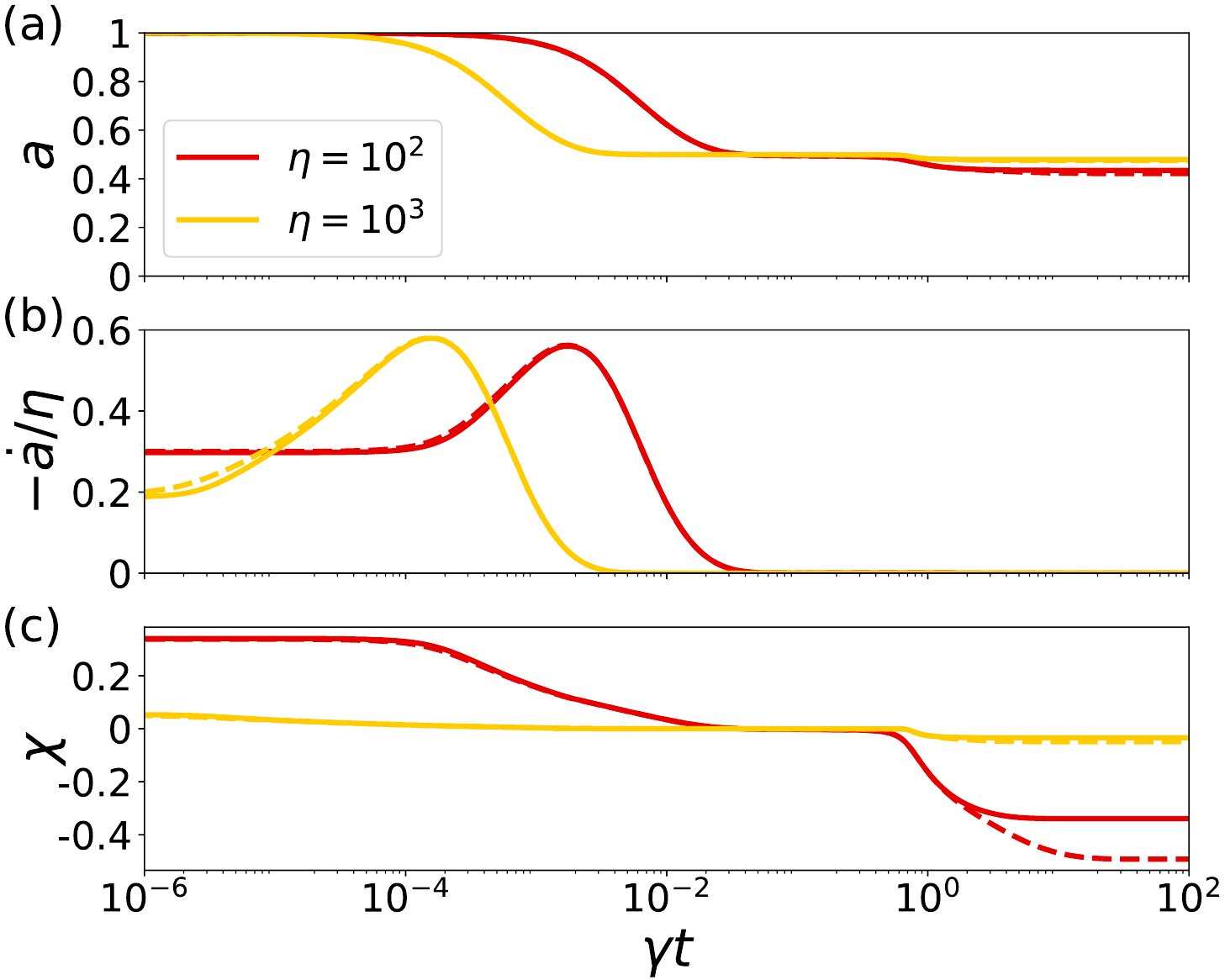}
\caption{To show the validity of the Taylor expansion of $q_0=k_0\sqrt{1 + i\chi}$, we compare the emission dynamics obtained with the exact expression (solid lines) and its Taylor-expanded form $q_0=k_0(1 + i\chi/2)$ (dashed lines).
(a) Upper-level population, (b) radiative intensity $\dot{a}/\eta$ and (c) expansion parameter $\chi$ in both scenarios and for $\eta=100$ ($C=\rho=10$) in red and $\eta=1000$ ($C=25$ and $\rho=40$) in yellow. The resulting dynamics are identical during the superradiant and subradiant phase, where $\chi$ remains small.}
\label{fig: time_evol_noTaylor}
\end{figure}

\begin{figure}
\includegraphics[width=\linewidth]{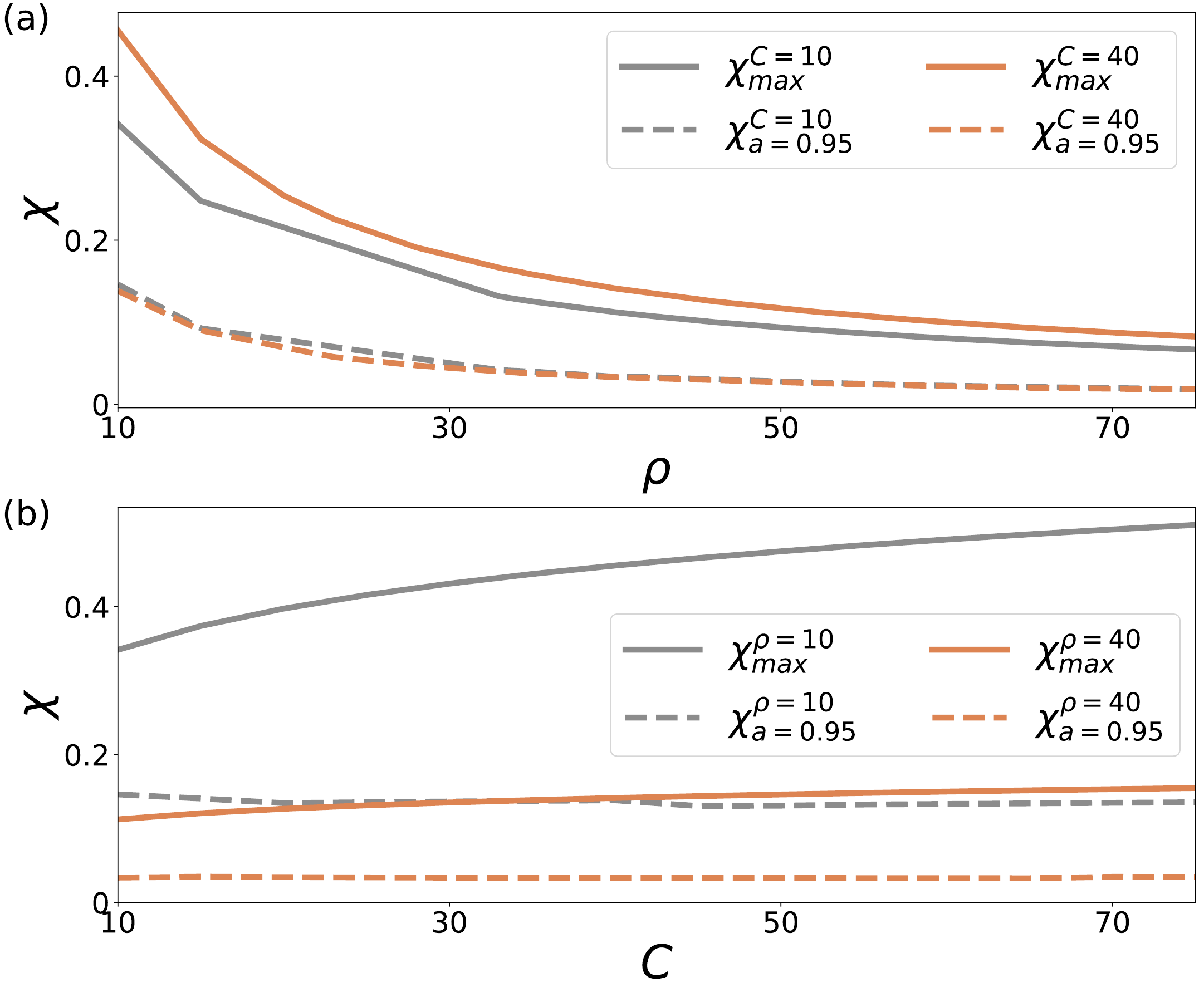}
\caption{Expansion parameter $\chi$ (a) for fixed $C=10$ (silver) and $C=40$ (orange) as a function of $\rho$, as well as (b) for fixed $\rho=10$ (silver) and $\rho=40$ (orange) as a function of $C$. Solid line shows the maximum achieved value of $\chi$ during the superradiant phase, which occurs at $t=0$, or equivalently $a=1$. Dashed lines show $\chi$ when $a=0.95$, and demonstrate that $\chi$ remains small during most of the superradiant decay for all $C$ and $\rho$.}
\label{fig: x_vsCrho}
\end{figure}

Let us focus on the non-driven case, for which

\begin{equation}
    \tilde{P}^{ret} = \frac{\wp^2}{\hbar^2} \mathcal{N} \frac{2a-1}{\gamma/2+\Gamma-i \Delta}
\end{equation}

Using the fact that $\gamma=\frac{\wp^2 k_0^3}{3 \pi \hbar \epsilon_0}$ and $C=\frac{\lambda_0^3 \mathcal{N}}{(2\pi)^2}=2\pi \frac{\mathcal{N}}{k_0^3}$, Eq. (\ref{eq: q0}) reduces to 

\begin{equation}
    q_0=k_0 \sqrt{1+i \gamma C \frac{2a-1}{\gamma/2+\Gamma-i \Delta}}
\end{equation}

One can define the quantity $\chi=\gamma C \frac{2a-1}{\gamma/2+\Gamma-i \Delta}$. For $|\chi| \ll 1$, $q_0$ can be Taylor expanded to first order

\begin{equation}
    q_0=k_0 \left( 1+i \gamma \frac{C}{2} \frac{2a-1}{\gamma/2+\Gamma-i \Delta} \right)
\end{equation}
\noindent and we retrieve the expression in \eq{eq:q0main} of the main text.
The expansion parameter $\chi$ therefore sets the error of the Taylor expansion.

This approximation holds in the subradiant regime, where $a \approx 0.5$ and $\Gamma \propto \eta = C \rho$ (see Fig. \ref{fig:4}), and $\chi$ therefore scales as $\chi \propto (2a-1)/\rho \rightarrow 0$ (region around $\gamma t = 10^{-1}$ in Fig. \ref{fig: time_evol_noTaylor}c). For the superradiant outburst, we initially have $\Gamma(t\rightarrow 0) \approx \exp{(\eta/\Gamma)}$. This requires a sublinear dependence of $\Gamma$ on $\eta$. As a result, $\chi$ scales inversely proportional to $\rho$ at early times and additionally exhibits a strong sublinear increase as a function of $C$, as depicted in Fig.~\ref{fig: x_vsCrho}a and b, respectively. While the initial values can reach $\chi>0.5$ for certain combinations of $C$ and $\rho$, $\chi$ rapidly decreases during the superradiant outburst due to the fast increase of $\Gamma$ and the simultaneous decrease of $2a-1$. As shown in Fig. \ref{fig: x_vsCrho}a-b, $\chi$ is always below $0.15$ when $a=0.95$ (dashed lines). Finally, $\chi$ becomes negative and grows again in absolute value during the radiation trapping regime ($a<0.5$). Although $\chi$ differs considerably for the exact and Taylor-expanded expressions ($\gamma t > 10^0$ in Fig. \ref{fig: time_evol_noTaylor}c), $\dot{a}$ is smaller than $10^{-5}$ in this regime and no significant difference is observed in terms of excited population $a$ (Fig. \ref{fig: time_evol_noTaylor}a).

In conclusion, $\chi$ is sufficiently small during most of the evolution and no significant difference is observed between the exact solution (solid lines) and the Taylor-expanded one (dashed lines) in Fig. \ref{fig: time_evol_noTaylor}a-b, which confirms the validity of the approximation.

\bibliography{sprd_ma}

\end{document}